\documentclass[journal,article,accept,pdftex,moreauthors]{Definitions/mdpi} 
\firstpage{1} 
\pubvolume{1}
\issuenum{1}
\articlenumber{0}
\pubyear{2022}
\copyrightyear{2022}
\datereceived{} 
\dateaccepted{} 
\datepublished{} 
\hreflink{https://doi.org/} % If needed use \linebreak
\pdfoutput=1

\usepackage{upgreek}

\def\solarm{\textit{M$_{\rm \odot}$}}
\def\ergs{\textit{$\rm erg\ s^{-1}$}}
\def\kms{\textit{$\rm km\ s^{-1}$}}
\def\whz{\textit{$\rm W\ Hz^{-1}$}}

\def\lx{\textit{$\rm L_{\rm X}$}}
\def\xlum{\textit{$\rm L_{2-10\,keV}$}}
\def\lrad{$\rm L_{1.4\,GHz}$}
\newcommand{\OIII}{[\rm{O\,\textsc{iii}}]}

\def\xmm{\textit{XMM-$Newton$}}
\def\swift{\textit{$Swift$}}
\def\chandra{\textit{$Chandra$}}

\Title{Exploring the Fundamental Mechanism in Driving Highest-velocity Ionized Outflows in Radio AGNs}

\TitleCitation{Title}

\Author{Ashraf Ayubinia $^{1,2,3}$*, Yongquan Xue $^{1,2}$*, Jong-Hak Woo $^{3}$, Huynh Anh Nguyen Le$^{1,2}$, Zhicheng He$^{1,2}$,Halime Miraghaei$^{4}$, Xiaozhi Lin$^{1,2}$}
\address{%
$^{1}$ \quad CAS Key Laboratory for Research in Galaxies and Cosmology, Department of Astronomy, University of Science and Technology of China, Hefei 230026, China\\
$^{2}$ \quad School of Astronomy and Space Science, University of Science and Technology of China, Hefei 230026, China \\
$^{3}$ \quad Department of Physics and Astronomy, Seoul National University, Seoul 08826, Republic of Korea \\
$^{4}$ \quad Research Institute for Astronomy and Astrophysics of Maragha (RIAAM), University of Maragheh, Maragheh, Iran}
\corres{ayubinia@mail.ustc.edu.cn, xuey@ustc.edu.cn}
\abstract{We investigate the ionized gas kinematics relationship with X-ray, radio and accreting properties using a sample of 348 nearby ($z < 0.4$) SDSS-FIRST-X-ray detected AGNs. X-ray properties of our sample are obtained from \xmm, \swift~and \chandra~observations.  We unveil the ionized gas outflows in our sample manifested by the non-gravitational broad component in \OIII $\uplambda$5007\AA~emission line profiles. From the comparison of the correlation of non-parametric outflow velocities (i.e., the velocity width, the maximal velocity of outflow and line dispersion) with X-ray luminosity and radio luminosity, we find that outflow velocities have similarly positive {correlations} with both X-ray and radio luminosity. After correcting for the gravitational component, we find that the \OIII~velocity dispersion normalized by stellar mass also increases with both X-ray luminosity and radio luminosity.  We  also find that for a given X-ray (radio) luminosity, radio (X-ray) luminous AGNs have higher outflow velocities than non-radio (non-X-ray) luminous AGNs. Therefore, we find no clear preference between X-ray luminosity and radio luminosity in driving high-velocity ionized outflows and conclude that both AGN activity and small-scale jets contribute comparably. Moreover, there is no evidence that our obscured AGNs are preferentially associated with higher velocity outflows. Finally, we find a turning point around log$(\uplambda_{Edd}) \simeq -1.3$ when we explore the dependency of outflow velocity on Eddington ratio. It can be interpreted considering the role of high radiation pressure (log$(\uplambda_{Edd})  \gtrsim -1.3$) in drastic reduction in the covering factor of the circumnuclear materials.}

\keyword{Active galactic nuclei; Supermassive black holes; Kinematics and dynamics kinematics and dynamics} 

\begin{document}
%=============================================================
%=============================================================
\section{Introduction} \label{sec:intro}
The coevolution of supermassive black holes and their host-galaxies is usually invoked to explain the well-known correlation of black hole mass (M$_{BH}$) with bulge mass {\cite{Kormendy2013, Son2022}} and stellar velocity dispersion of the host-galaxy \cite{Gultekin2009,Woo2013}. These black holes grow through the feeding on circumnuclear materials and are visible as active galactic nuclei (AGNs). The outgoing emission spanning the entire electromagnetic spectrum is the product of accretion. The energy released effectively couples to the material in the host-galaxy and impacts on the environment of the black hole (namely, AGN feedback), from small to large scales. However, it is still not well understood {how} AGN feedback affects the interstellar medium (ISM) of the host-galaxy and more dedicated models and observations are required \cite{Gaspari2020, Woo2017, Woo2020, Harrison2017}.\ 

Multi-phase outflows, frequently detected in local and high redshift galaxies, are capable of interacting with ISM and hence, they are one of the fundamental tracers of AGN feedback (see Fabian \cite{Fabian2012} for a review). A wide range of studies invoked the negative feedback scenario to explain the suppression of star formation (SF) by AGN-driven outflows in which powerful outflows heat \cite{Shangguan2018} or sweep up gas reservoirs from the host-galaxy \cite{Feruglio2015} and reduce SF activity \cite{Chen2022}. However, some studies found evidence of positive feedback in which AGN-driven outflows trigger local SF through the compression of ISM \cite{Cresci2015, Shin2019} and/or providing physical conditions for SF occurring within outflows \cite{Maiolino2017}. {Most recently, Bessiere and Almeida \cite{Bessiere2022} put up a scenario in which quasar-driven outflows can simultaneously produce both modes of feedback.} \

There is no clear mechanism for driving galactic-scale outflows. In luminous AGNs accreting close to Eddington limit, it is believed that the strong radiation pressure pushes the gas away from the nuclear region and forms the observed outflows (so-called radiative or quasar-mode feedback). However, an alternative feedback mechanism, so-called radio or kinetic mode feedback, is in favour of radio galaxies with relativistic jets usually hosted by massive elliptical galaxies. According to this scenario, low-luminosity AGNs are radiatively inefficient to expel gas to large scales, instead, radio jets that carry much of the energy output in form of kinetic energy, can be {effective} enough to impact on ISM and drive outflows. Particularly, young radio jets with low to moderate radio power (\lrad~$\lesssim 10^{25}$ \whz) can penetrate into ISM and {alter} gas kinematics. Higher power radio jets are presumed to be highly collimated that pierce the ISM without disturbing gas significantly (see the discussion in Wylezalek and Morganti \cite{Wylezalek2018}). However, the fact is that the observational disentanglement of different modes of AGN feedback is not straightforward.\

Recent observational studies have tried to shed light on the link between outflows and radio emission \cite{Zakamska2016, Jarvis2019, Singha2021}. However the results are not unanimous. Mullaney et al. \cite{Mullaney2013} analyzed the \OIII$\lambda$5007\AA~profile for a large sample of nearby AGNs and found that AGNs with \lrad~$> 10^{23}$ \whz~are much more likely to exhibit extremely broad \OIII$\lambda$5007\AA~lines compared to weak radio AGNs. They concluded that in comparison with \OIII$\lambda$5007\AA~luminosity, radio luminosity has the most profound effect on the disturbing gas kinematics in radio AGNs. Moreover, recent integral field spectroscopy observations found evidence of enhanced line widths in AGNs that host low-power jets with low inclination with respect to the galaxy disc, suggestive of strong interaction between radio jets and the ISM \cite{Venturi2021}. In contrast, using a large sample of optically-selected Type 2 AGNs, Woo et al. \cite{Woo2016} reported that while the kinematics of ionized gas traced by \OIII$\lambda$5007\AA~emission line strongly correlates with AGN radiation, ionized gas outflows are not connected to the radio activity. A similar conclusion was given for Type 1 AGNs \cite{Rakshit2018}. It is worth mentioning that both Woo et al. \cite{Woo2016} and Rakshit and Woo \cite{Rakshit2018} also observed that the width of \OIII$\lambda$5007\AA~profiles (\OIII~velocity dispersion) increases with radio luminosity, however, they noticed such a trend fades after correcting for the effect of host-galaxy gravitational component (see Section \ref{sec:radio_properties}). \

{Since the radio luminous sources are predominantly associated with AGNs with higher X-ray luminosity, it is important to control X-ray luminosity when we explore the correlation between ionized gas kinematics and radio emission. Despite the great insight provided by previous works using large samples, those studies suffer from the lack of such controlling. A recent study by Harrison et al. \cite{Harrison2016} provided a sample of X-ray detected radio AGNs with ionized outflow signature. When compared to non-radio-luminous sources, they found that a higher fraction of radio-luminous AGNs have high velocity line widths. However, due to lack of enough targets, they were unable to test their results while X-ray luminosity is controlled. Hence, in this work our primary interest is to explore the dependency of the kinematic properties of ionized gas outflows on both X-ray luminosity and radio luminosity, while we control luminosities during our investigations to understand which mechanism is more responsible for driving highest-velocity ionized outflows using a sample of nearby AGNs.} In Section \ref{sec:sample} we describe our sample construction. In Section \ref{sec:data_analysis} we present our data analyses. In Section \ref{sec:discussion} we discuss the relation of ionized gas outflows with X-ray, radio and accreting properties of our sample and in Section \ref{sec:conclusions} we draw our conclusions. A cosmology with $H_{0}$ = 71 \kms~Mpc$^{-1}$, $\Omega_{m}$ = 0.27 and $\Omega_{\Lambda}$ = 0.73 is used throughout the paper.

%=============================================================
%=============================================================

\section{Sample Construction} \label{sec:sample}
\begin{figure}
\centering
\includegraphics[width=.55\linewidth]{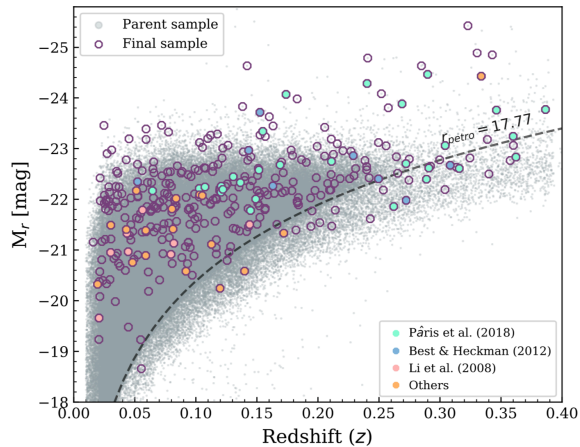}
\caption{Absolute extinction-corrected r-band magnitude as a function of redshift. Grey dots and purple circles show our parent sample and final sample; the sources identified as AGN in some other works are also shown (Section \ref{sec:SFs}). The dashed curve shows the SDSS Petrosian r-band magnitude corresponding of $r_{petro}$ = 17.77.}
\label{fig:magnitude-z}
\end{figure}

Our initial sources are selected from the Sloan Digital Sky Survey (SDSS) which is a large-area multiband spectral imaging and spectroscopic redshift survey and covers 14,555 square degrees of the sky. We use the data from the Data Release 15 (DR15 \cite{Aguado2019}) as the third data release of the fourth phase of the SDSS (SDSS-IV) and that includes all the sky coverage of prior releases. By applying the Structured Query Language (SQL), we collect emission-line objects that satisfy the following:
\begin{itemize}
    \item Being spectroscopically classified as ``GALAXY'' or ``QSO'' by the SDSS DR15.
    \item Having a secure redshift of $0.01 < z < 0.4$ (\textit{ZWARNING} = 0, indicating a confident spectroscopic classification). {Galaxies at $z < 0.01$ are usually very extended, resulting in a lack of a reliable optical position. On the other hand, at higher redshifts of $z > 0.4$ the quality of the SDSS spectrum is usually low.}
    \item Having a median signal-to-noise ratio (S/N)  $\geq$ 10.
    \item Having four emission lines, H$\alpha$, H$\beta$, \OIII $\lambda 5007$ and [N II]$\lambda 6584$, whose intensities are at least 2 times larger than their corresponding measured errors.
\end{itemize}
Using the above criteria we obtain 322,006 emission-line galaxies and hereafter we define them as our parent sample. The grey points in Figure \ref{fig:magnitude-z} represent the absolute extinction-corrected r-band magnitudes of our parent sample as a function of redshift. The r-band magnitude is corrected using the dust map of Schlafly and Finkbeiner \cite{Schlafly2011} and extinction law of Fitzpatrick \cite{Fitzpatrick1999}.

%===========
%===========

\subsection{Cross-correlation: SDSS-FIRST} \label{sec:radio}
{Since the area of the Faint Images of the Radio Sky at Twenty-centimeters (FIRST; \cite{Becker1995}) survey has been chosen to coincide with that of SDSS, nearly one third of its sources have optical counterparts in the SDSS.  Therefore, we utilize the FIRST survey to find our radio sources.} The survey uses the NRAO Very Large Array (VLA) to observe 9,000 square degrees in the north Galactic cap and a smaller strip along the celestial equator at 1365 and 1435 MHz. To select galaxies with detected radio core emission (or small-scale jet), we cross-correlate our parent sample with all FIRST sources. It is important to reduce the number of coincidental false matches during the cross-correlation process.  Therefore, we test three different matching radii and estimate the number of random false matches in each case. At first, we search for the best matched (nearest-neighbour) radio sources within 2$^{\prime\prime}$ angular separation from the optical core and we find 17,918 matches. We next offset the positions (RA and DEC) of the SDSS sources by 1$^{\circ}$ and re-matched them with the FIRST catalogue. We find on average 28 random matches, corresponding to $\sim0.2\%$ contamination from coincidental matches ($>$ 99.8\% efficiency). We then repeat this algorithm for matching radius 3$^{\prime\prime}$ and find that the selection efficiency is $\sim99.6\%$ (with on average 68 random matches among 18,372 matches). The fraction of false matches increases to $\sim1\%$ (efficiency $\sim99\%$) when we cross-correlate SDSS galaxies with the FIRST sources within 5$^{\prime\prime}$  angular separation. Considering the fact that none of the false matches is included in our final sample when we use the matching radius of either 2$^{\prime\prime}$ or 3$^{\prime\prime}$, we finally adopt a match criterion of 3$^{\prime\prime}$ for cross-correlating our parent sample with the FIRST sources to have a more complete sample. We classify these matched sources into resolved and unresolved radio galaxies based on the definition from Banfield et al. \cite{Banfield2015} as a resolved source satisfies the criterion:

\begin{equation}
\rm {S}_{\rm peak}/{S}_{\rm int} < 1.0 - (0.1 / log({S}_{\rm peak})), \\
\end{equation}
 
where ${S}_{\rm peak}$ and ${S}_{\rm int} $ are the peak flux density and integrated flux density measured in mJy at 1.4 GHz, respectively. We next exclude sources with the radio flux (integrated flux for resolved and peak flux for unresolved sources; see also Zhu et al. \cite{Zhu2020}) below the FIRST detection limit (5$\rm$ $\sigma$ flux limit is 1 mJy). Therefore, we consider the surviving 18,153 SDSS-FIRST detected galaxies as the `radio sample' for the following investigations. \\

%===========
%===========

\subsection{Cross-correlation: SDSS-FIRST-X-ray Surveys} \label{sec:x-ray}
We identify X-ray counterparts of our radio sample {within 5$^{\prime\prime}$ from the optical core using three catalogs from the sensitive, wide and flexible X-ray surveys as follows}. I) The 3XMM-DR8 \cite{Rosen2016} which is the third-generation catalogue of serendipitous X-ray sources from the \xmm~observatory and contains 531,454 unique X-ray sources drawn from a total of 10,242 \xmm~EPIC observations. A number of 522 of our galaxies in the radio sample have X-ray counterparts in this catalogue.  II) The 1SWXRT catalogue of the serendipitous sources detected in the 7 years (2005-2011) of \swift-XRT observations with exposure time longer than 500 s \cite{D'Elia2013} including a total of $\sim$ 36,000 distinct X-ray sources. We find that 227 sources of our radio sample are included in this catalogue. III) The full z $<$ 0.4 spectroscopic \chandra/SDSS catalogue \cite{Trichas2013} including 617 sources from the \chandra~Source catalogue matched to the SDSS DR7 spectroscopic catalogue. A cross-matching procedure of our radio sample with this catalogue returns 117 sources. There are 106 radio galaxies which are common between at least two X-ray catalogs. In this case, we choose the survey with the highest X-ray photon count. We also estimate the false-match rates in cross correlating of radio sample with X-ray catalogs similar as in Section \ref{sec:radio} for matching radii 5$^{\prime\prime}$. Except two random matches from \swift~catalog, we find no false-match cases. Since these two sources do not impact on our results we keep them. As such, we obtain X-ray data for 760 individual sources from our radio sample. Afterward, we call these SDSS-FIRST-X-ray detected galaxies as the `full sample'.  \\

%===========
%===========

\subsection{Exclusion of Star-forming Galaxies} \label{sec:SFs}
\begin{figure}
\centering
\includegraphics[width=.55\linewidth]{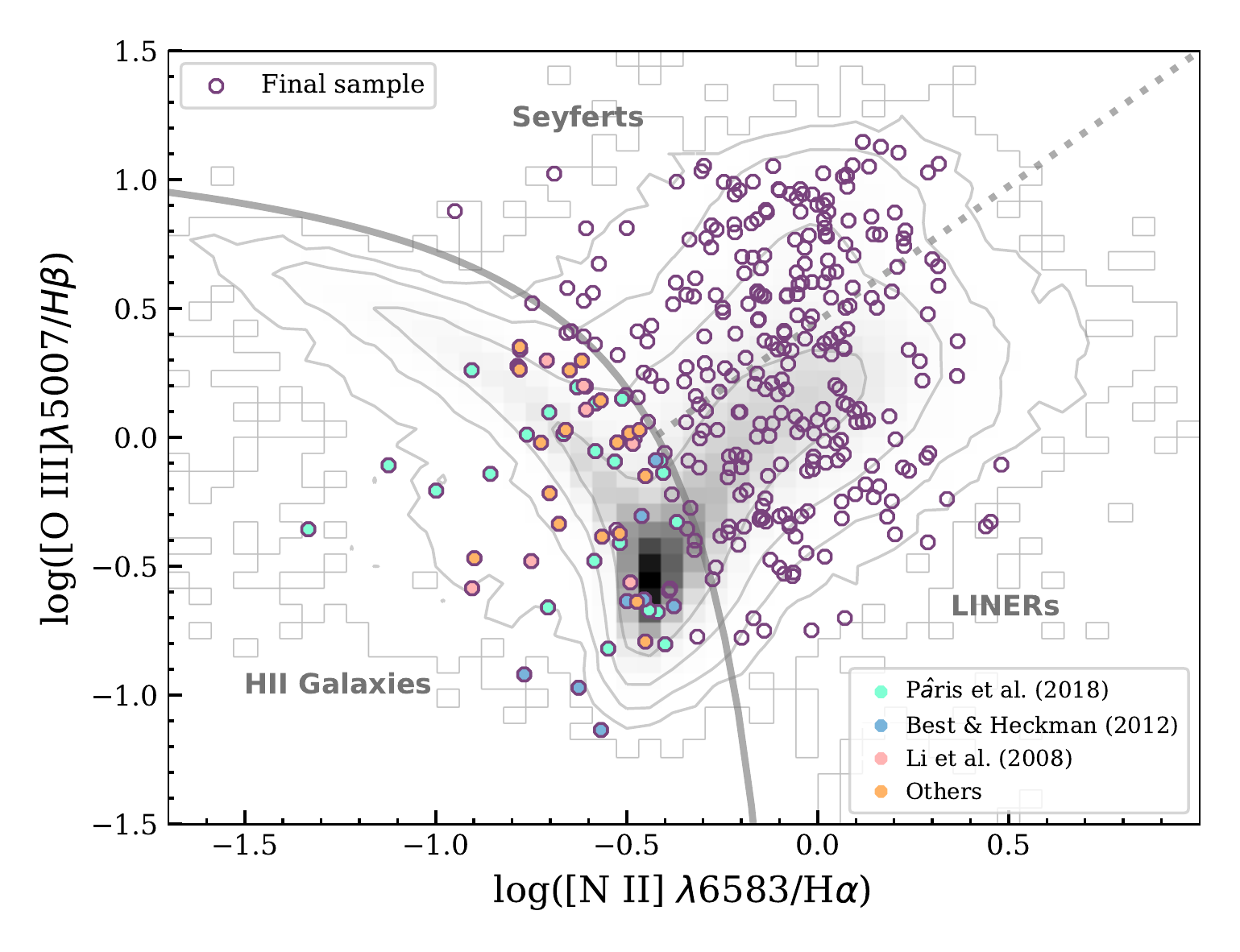}
\caption{‌The BPT diagram with our final sample (denoted by purple circles). The solid curve is the classification curve from Kauffmann et al. \cite{Kauffmann2003} and the dotted line shows the division of Seyferts and LINERs (Low Ionization Nuclear Emission-Line Regions) from Schawinski et al. \cite{Schawinski2007}. The contours show our parent sample.}
\label{fig:bpt}
\end{figure}

It is important to avoid selecting pure star-forming galaxies whose X-ray emission is dominated by star formation. Here, we explain our efforts to discriminate between star-forming galaxies and AGNs among our full sample. Given that a pure star-forming galaxy should have an extreme star-formation rate (SFR $\gtrsim$~1900 \solarm~yr$^{-1}$) in order to produce an X-ray luminosity \lx~$\sim$~10$^{42}$ \ergs~\cite{Harrison2016}, we select sources with hard X-ray luminosity of \xlum~$> 10^{42}$~\ergs, where X-ray emission is AGN-dominated. For this step, we estimate X-ray luminosity using the observed fluxes from X-ray catalogues with the assumption of photon index $\Gamma$ = 1.7 for all sources (see Section \ref{sec:x-ray}). The selection of other values for photon index does not impact on our final sample significantly. This process reduces our full sample into 348 sources. We next plot these remaining targets on the BPT diagram \cite{Baldwin1981} in Figure \ref{fig:bpt}. A small fraction of our sources (70 targets) reside in the star-forming region defined by Kauffmann et al. \cite{Kauffmann2003}. It is very possible that radio-quiet AGNs (whose radio emission is produced by a variety of possible mechanisms such as star formation (see Panessa et al. \cite{Panessa2019} for a review) overlap star-forming region based on emission-line diagnostic. Therefore, we check the reliable classification for each source in the literatures when it is available. We summarize this procedure as follows:

- We find that 28 targets are identified as quasar in the SDSS DR14 quasar catalogue (DR14Q \cite{Paris2018}).

- We find 16 targets labeled as radio AGN by either Best and Heckman \cite{Best2012} or Li et al. \cite{Li2008} who provided radio AGN catalogues in total including $\sim$~19,000 sources.

- Finally, 18 targets are identified as AGNs in these studies: Anderson et al. \cite{Anderson2007}; Sun and Shen \cite{Sun2015}; V\'eron-Cetty and V\'ron \cite{Cetty2010}; Monroe et al. \cite{Monroe2016}.

Overall, a very high fraction (i.e., 62/70) of the sources which reside in the BPT star-forming region are classified as AGNs in previous studies. Therefore, we keep all the 70 targets and identify 348 AGNs in total detected in optical, radio and X-ray as our `final sample' (see Figure \ref{fig:magnitude-z} for the redshift distribution of absolute extinction-corrected r-band magnitude for our final sample). 

\section{Data Analyses} \label{sec:data_analysis}

%===========
%===========

\subsection{Detection of Ionized Outflows} \label{sec:badass}
\begin{figure}
\centering
\includegraphics[width=.85\linewidth]{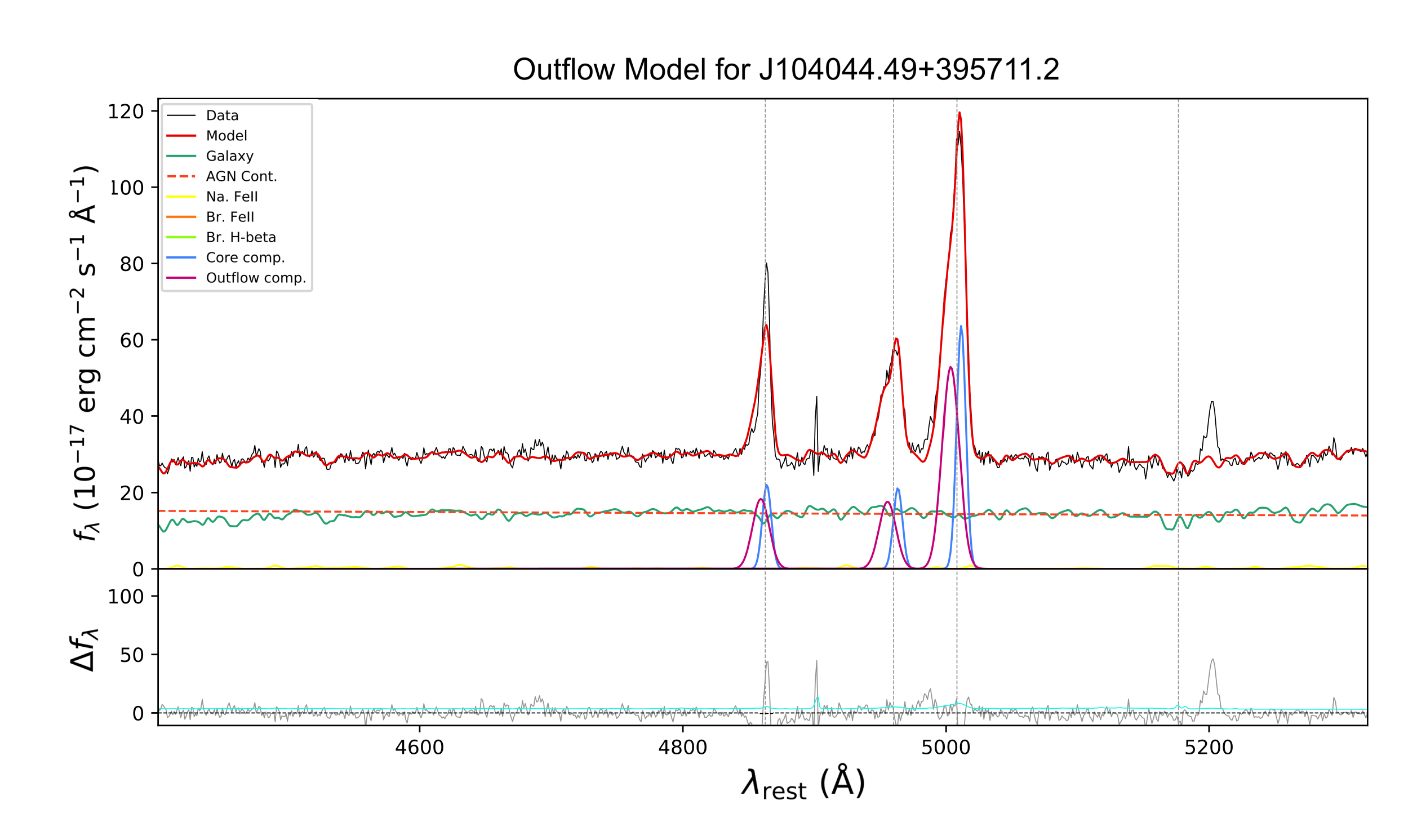}
\caption{‌An example of H$\beta$+\OIII$\lambda\lambda$4959,5007\AA~region fitted by BADASS. The spectrum and the overall best-fit model are presented in black and red, respectively. The individual fitting components are designated by different colors.}
\label{fig:opt_spectra}
\end{figure}

In order to unveil the non-gravitational ionized gas outflows and constrain their kinematic characteristics in our sample, we utilize the \OIII$\lambda$5007\AA~(hereafter \OIII) emission line. We employ the Bayesian AGN Decomposition Analysis for SDSS Spectra (BADASS; \cite{Sexton2020}) package specialized for fitting AGN spectra through the deconvolution of AGN and host-galaxy spectra. BADASS includes all spectral components such as the power-law continuum, the FeII pseudo-continuum, stellar line-of-sight velocity distribution and  individual spectral line components. We fit each spectrum using single-Gaussian (no-outflow) and double-Gaussian (outflow) models in the wavelength range of 4400-5800 \AA~(see Section 2.2 in Sexton et al. \cite{Sexton2020} for fitting procedure). We adopt the non-gravitational component in \OIII~when the statistical F-test suggests a double-Gaussian model with more than 95\% confidence level in comparison to the single-Gaussian model. Besides, we visually check the quality of fit to exclude sources with poor fit. Figure \ref{fig:opt_spectra} shows an example of spectrum illustrating the best-fit model in the H$\beta$+\OIII$\lambda$$\lambda$4959,5007\AA~region.
By default, BADASS utilizes the Markov-Chain Monte Carlo (MCMC) approach to estimate parameters and associated uncertainties. Some relevant quantities obtained from spectral fitting are AGN luminosity at 5100 \AA~($\lambda L_{\text{5100\AA}}$), stellar velocity dispersion ($\upsigma_{*}$), core (outflow) velocity dispersion and core (outflow) velocity offset from the rest-frame wavelength of the line. We also estimate the maximal velocity of the outflows by the relation from Harrison et al. \cite{Harrison2014}:

\begin{align}\label{eq:v_max}
  \rm V_{max}&= \Delta V + \frac{W_{80}}{2},
\end{align}

where $\Delta$V is the velocity offset of the outflow component with respect to the velocity offset of the core component, and $\rm W_{80}=1.09 \times \;\rm{FWHM}$ (FWHM: the full width at half maximum) is the velocity width of the line that contains 80\% of the outflow flux.

\begin{figure}
\centering
\includegraphics[width=0.55\linewidth]{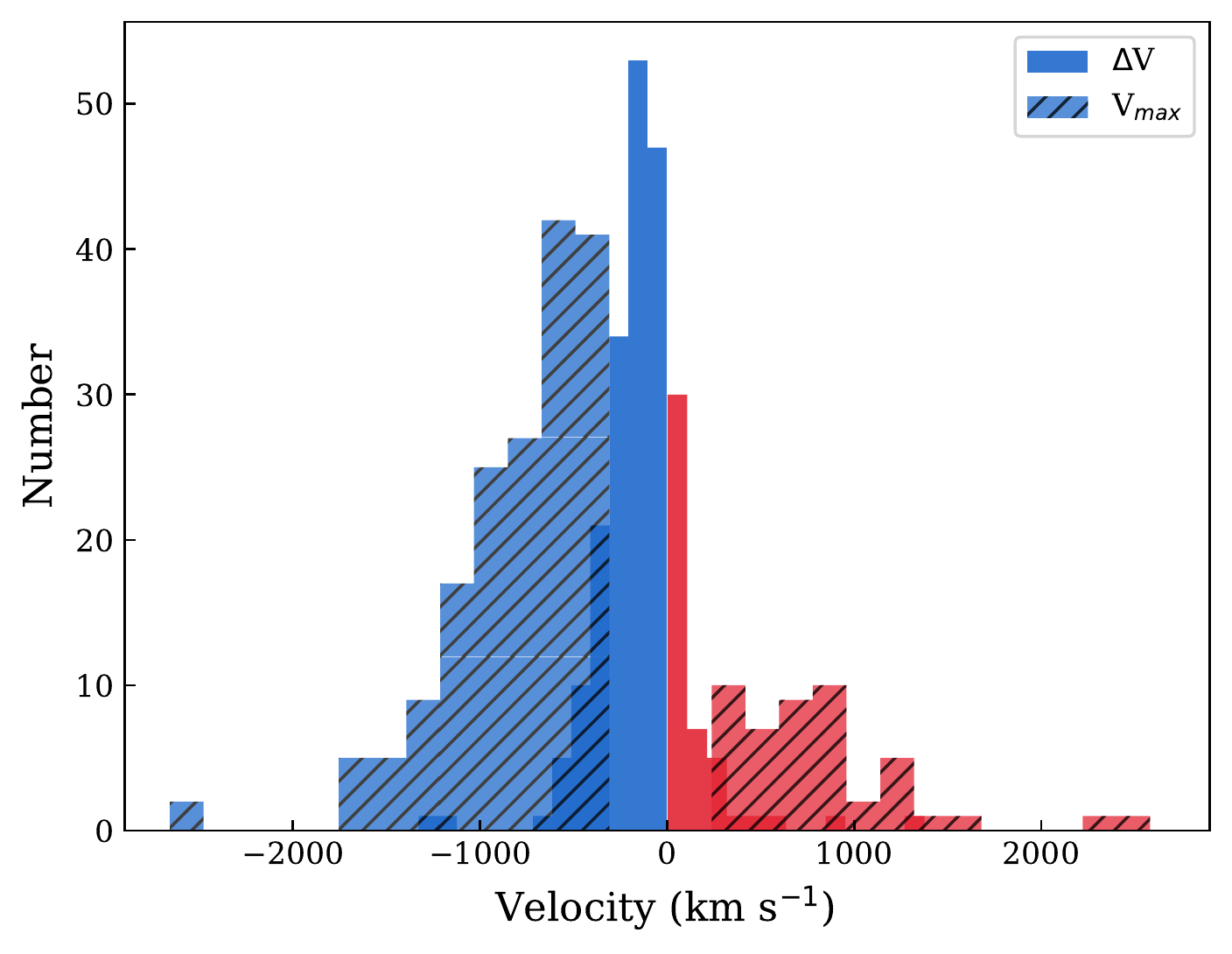}
\caption{‌The number distributions of $\Delta$V and $\rm V_{max}$ for our final sample with outflow signatures. The approaching and receding outflows have been showed in blue and red, respectively.}
\label{fig:velocity_hist}
\end{figure}

Based on our analysis described above, 63\% (220/348) of AGNs in our final sample exhibit ionized outflow features in their optical spectra. Using the measurements of the \OIII~emission-line profiles provided by Mullaney et al. \cite{Mullaney2013}, Molyneux et al. \cite{Molyneux2019} also reported that $\sim$ 77\% of their selected radio AGNs require the second broader component in \OIII~emission line profile (indicative of outflows). These high detection rates suggest the high prevalence of ionized outflows in radio AGNs. As shown in Figure \ref{fig:velocity_hist}, the majority of outflows in our sample are blueshifted with respect the core component of the line (see also Woo et al. \cite{Woo2016} for non-radio AGNs).  

%===========
%===========

\subsection{X-ray Data Reduction and Spectral Fitting} \label{sec:x-ray-data}
X-ray data of our sample is based on observations obtained with \xmm, \swift~and  \chandra.\

The \xmm~data reduction is carried out using the \xmm~Scientific Analysis Software (SAS; v.16.1.0) package with the most recent calibration files. We process the EPIC (European Photon Imaging Camera) data with $epchain$ and $emchain$ tasks and produce light curves and images with events selected by patterns corresponding to single and double pixel events for the EPIC PN camera (0 and 1–4), and patterns 0–12 for the EPIC MOS cameras. We also clean light curves for large background flares. The source spectrum is extracted from a source-centered region with a radius selected properly. The background spectrum is taken from a source-free region (with a larger radius) on the same instrument CCD as the source. For very bright sources, we check the pile-up effect and remove it when it is evident. Our priority is to analyse data from PN detector due to its larger effective area; however, in the case that the source rides on the gap between two CCD chips, we use data from MOS cameras. Therefore, we obtain 0.3-10 keV spectra corresponding to a single instrument for 94/220 AGNs with detected ionized outflows.\

The \swift/XRT data processing is performed using the standard data pipeline package (\texttt{xrtpipeline}; v.0.13.3). We only utilize photon-counting (PC) mode data {which retain 2-dimensional imaging and full-energy resolution}, and select the standard grade filtering (0–12 for PC) according to the XRT nomenclature. The source and background spectra are extracted using the same strategy we consider for \xmm~observations. When the XRT data are affected by pile-up effect (i.e., source count rate $\gtrsim$ 0.6), we correct it using the model described in Evans et al. \cite{Evans2020}. Finally, we obtain 0.3-10 keV spectra for 100/220 AGNs from \swift/XRT observations.\

\begin{figure}
\centering
\includegraphics[width=.6\linewidth]{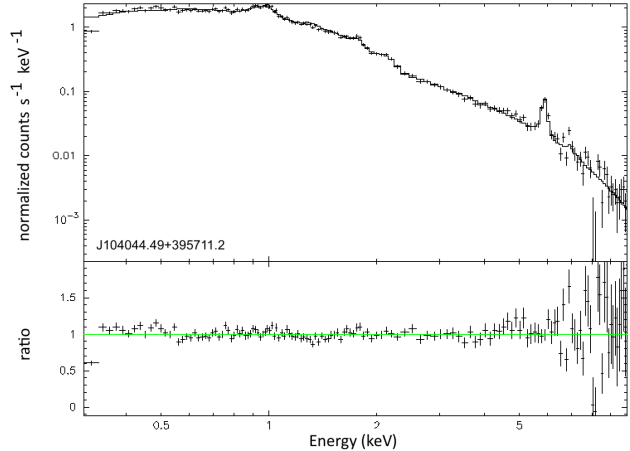}
\caption{‌An example of X-ray spectral fitting of our sample in the range of 0.3-10 keV from \xmm~observations with the corresponding optical spectrum presented in Figure \ref{fig:opt_spectra}.}
\label{fig:xray_spectra}
\end{figure}

To analyse X-ray spectra produced above, we bin each spectrum with \texttt{GRPPHA} to contain at least one count in each bin so that the Cash statistics \cite{Cash1979} can be applied which has been proven to work well in the case of low-count spectra. We adopt a threshold in source net count (background-subtracted) more than 40 to ensure the reliability of the spectral-fitting results. Therefore, we exclude 37 sources with net count $<$ 40. Spectral fitting is performed with XSPEC \cite{Arnaud1996} (v.12.9.1) in the range of 0.3 to 10 keV. In the first step, we fit each spectrum with the simple absorbed power-law model with fixed Galactic absorption as the baseline model. The values of Galactic absorption column density have been taken from the HI4PI survey \cite{HI4PI2016}. This model is sufficient to describe the observed X-ray spectrum satisfactorily for the majority ($\sim$ 66\%) of our AGNs. When the soft excess is prominent in the residual, we add an additional component; either a second power-law model (\texttt{pow}), thermal absorption (\texttt{mekal}) or black-body emission (\texttt{bb}), to the baseline model depending on which component is physically more motivated and optimise the spectral fitting. Moreover, when a significant iron K$\alpha$ fluorescent emission line is found (F-test probability $>$ 99\%), we fit this line with a Gaussian model (\texttt{zgauss}) as the energy of line is fixed at 6.4 keV and the width of line is fixed at  $\sigma$ = 50 eV. To measure the rest-frame absorption-corrected 2-10 keV luminosity, we use the \texttt{clumin} command after removing absorption components (\texttt{wabs*zwabs}) and fixing all parameters into the best-fit values. As an example, Figure \ref{fig:xray_spectra} shows the X-ray spectrum of a source from our sample whose optical spectrum is presented in Figure \ref{fig:opt_spectra}. \

It should be noted that for 26 \chandra-detected AGNs in our sample, we do not produce X-ray spectra and we use the results provided by Trichas et al. \cite{Trichas2013}. Briefly, they fixed Galactic absorption column density at 21 cm and fitted three power-law models (i.e., \texttt{PL} with no intrinsic absorption component, \texttt{PLfix} with a fixed $\Gamma$ = 1.8 and \texttt{PL-abs} while both $\Gamma$ and column density are free) for each source using Cash statistics. We obtain 2-10 keV X-ray luminosity converted from the 2-8 keV luminosity measurements provided by Trichas et al. \cite{Trichas2013} (see their Table 1 for details) assuming $\Gamma$ = 1.7.  \

%===========
%===========
\subsection{Other Parameters} \label{sec:others}
The majority of our targets have FWHM(H$\beta$) $>$ 1000 \kms~ which can be classified as Type 1 AGNs. Therefore, we adopt the virial relation given in Woo et al. \cite{Woo2015} to calculate black hole mass based on the FWHM of the H$\beta$ line and the luminosity at 5100\AA as

\begin{equation}\label{equ:bh_mass1}
    \frac{M_{\mathrm{BH}}}{M_{\mathrm{\odot}}} = 1.12 \times 10^{6.819} (\frac{FWHM_{\mathrm{H\upbeta}}}{ 10^{3}\ erg\ s^{-1}})^{2} (\frac{\uplambda L_{5100}}{10^{44}\ erg\ s^{-1}})^{0.533}
\end{equation}

The black hole masses cover the interval of $\rm 5.1 \lesssim log(M_{BH}/M_{\odot}) \lesssim 9.2$. Next we estimate Eddington luminosity ($\rm L_{Edd} = 1.26 \times 10^{38}\ M_{BH}$). We derive the bolometric luminosity using $\rm L_{bol} = 9 \times L_{\text{5100\AA}}$  \cite{Kaspi2000}. Finally, the Eddington ratio is determined using the relation $\rm \uplambda_{Edd} = L_{Bol}/L_{Edd}$.

%@*@*@*@*@*@*@*@*@*@*@*@*@*@*@*@*@*@*@*@*@*@*@*@*@*@*@*@*@*@*@*@*@*@*@*@*@*@*
%@*@*@*@*@*@*@*@*@*@*@*@*@*@*@*@*@*@*@*@*@*@*@*@*@*@*@*@*@*@*@*@*@*@*@*@*@*@*
\begin{table}[h!]
  \begin{center}
  \setlength{\tabcolsep}{1.2pt}
   \scriptsize
    \caption{Main parameters of our sample (183 targets) obtained from optical and X-ray analysis.}
    \label{tab:parameters}
    \begin{tabular}{cccccccccccc} % <-- Alignments: 1st column left, 2nd middle and 3rd right, with vertical lines in between
    \hline\hline
     \text{ID} & \text{Redshift} & \text{log \lrad}  & \text{log(\xlum)}  &     \text{log$\rm(N_{H})$}  & \text{log($\lambda L_{\text{5100\AA}}$)} & \text{log($\rm L_{\OIII}$)}  & \text{W80}  &
	 \text{$|$V$_{\rm{max}}|$}  & \text{$\sigma_{\OIII}$}  &   \text{log($\rm M_{BH}$)}  & \text{$\lambda_{Edd}$}\\
	 \\
       \text{}   &  \text{}   &  \text{[\ergs]}  &  \text{[\ergs]}  &  \text{$[\textrm{cm}^{\textrm{--2}}\textrm{]}$}  &  \text{[\ergs]}  &  \text{[\ergs]}  &  \text{[\kms]}  &   \text{[\kms]}  &  \text{[\kms]}  &
         \text{[$\rm M_{\odot}$]    }  &  \text{}\\
         \\
                    \text{(1)}  &\text{(2)}  & \text{(3)} & \text{(4)} & \text{(5)} & \text{(6)}  & \text{(7)} & \text{(8)} & \text{(9)}  & \text{(10)}  & \text{(11)}  & \text{(12)}\\     
      \hline
      \\
J000202.95-103038.0 & 0.103 & 38.72 & 43.04$^{+ 0.05}_{- 0.05}$ & 20.00$^{+ 0.79}$ & 43.08$^{+ 0.03}_{- 0.03}$ & 41.38$^{+ 0.02}_{- 0.02}$ & 709.38$^{+ 27}_{- 25}$ & 495.11$^{+ 20}_{- 21}$ & 207.62 & 7.45 & 0.031\\
J001047.89+145101.3 & 0.042 & 38.21 & 42.38$^{+ 0.05}_{- 0.04}$ & 23.05$^{+ 0.13}_{- 0.17}$ & 40.67$^{+ 0.29}_{- 0.26}$ & 40.35$^{+ 0.02}_{- 0.03}$ & 880.05$^{+ 92}_{- 94}$ & 687.84$^{+ 93}_{- 80}$ & 190.26 & 5.87 & 0.005\\
J001137.24+144201.3 & 0.132 & 39.00 & 43.21$^{+ 0.01}_{- 0.01}$ & 20.40$^{+ 0.20}_{- 0.38}$ & 43.29$^{+ 0.03}_{- 0.02}$ & 41.35$^{+ 0.01}_{- 0.01}$ & 1870.53$^{+ 55}_{- 55}$ & 1164.95$^{+ 43}_{- 46}$ & 580.91 & 7.64 & 0.032\\
J014816.25+001944.9 & 0.092 & 41.01 & 42.36$^{+ 0.10}_{- 0.11}$ & 20.00$^{+ 1.1}$ & 42.29$^{+ 0.15}_{- 0.12}$ & 41.48$^{+ 0.03}_{- 0.03}$ & 712.94$^{+ 28}_{- 27}$ & 448.18$^{+ 19}_{- 20}$ & 204.75 & 6.17 & 0.094\\
J020615.98-001729.2 & 0.043 & 38.24 & 43.73$^{+ 0.002}_{- 0.002}$ & 20.43$^{+ 0.29}$ & 43.42$^{+ 0.01}_{- 0.01}$ & 41.27$^{+ 0.01}_{- 0.01}$ & 1626.62$^{+ 349}_{- 413}$ & 982.87$^{+ 209}_{- 244}$ & 338.89 & 7.97 & 0.020\\
J022014.58-072859.2 & 0.213 & 39.28 & 44.39$^{+ 0.02}_{- 0.02}$ & 20.88$^{+ 0.14}_{- 0.20}$ & 44.05$^{+ 0.01}_{- 0.01}$ & 41.72$^{+ 0.02}_{- 0.02}$ & 1533.36$^{+ 124}_{- 113}$ & 925.37$^{+ 74}_{- 70}$ & 417.13 & 8.76 & 0.014\\
J030117.99-003842.3 & 0.183 & 39.57 & 42.40$^{+ 0.04}_{- 0.05}$ & ...   & 41.97$^{+ 0.31}_{- 0.30}$ & 41.60$^{+ 0.02}_{- 0.02}$ & 1070.52$^{+ 57}_{- 49}$ & 635.17$^{+ 34}_{- 31}$ & 335.14 & 7.29 & 0.003\\
J030639.57+000343.1 & 0.107 & 39.11 & 43.39$^{+ 0.05}_{- 0.01}$ & 20.00$^{+ 0.11}$ & 43.72$^{+ 0.01}_{- 0.01}$ & 41.78$^{+ 0.01}_{- 0.01}$ & 1093.38$^{+ 32}_{- 29}$ & 674.58$^{+ 19}_{- 18}$ & 317.57 & 7.52 & 0.114\\
J032525.36-060837.8 & 0.034 & 38.82 & 42.35$^{+ 0.01}_{- 0.01}$ & 20.00$^{+ 0.06}$ & 42.8$^{+ 0.04}_{- 0.04}$ & 41.51$^{+ 0.01}_{- 0.01}$ & 927.77$^{+ 8}_{- 9}$ & 665.50$^{+ 7}_{- 7}$ & 279.02 & 8.20 & 0.003\\
J073623.13+392617.7 & 0.118 & 39.22 & 44.23$^{+ 0.06}_{- 0.06}$ & 20.00$^{+ 0.58}$ & 44.29$^{+ 0.004}_{- 0.004}$ & 42.28$^{+ 0.09}_{- 0.09}$ & 1564.49$^{+ 26}_{- 26}$ & 815.21$^{+ 16}_{- 16}$ & 448.75 & 8.27 & 0.075\\
J074738.39+245637.3 & 0.130 & 39.16 & 43.82$^{+ 0.04}_{- 0.04}$ & 20.00$^{+ 0.67}$ & 43.92$^{+ 0.01}_{- 0.01}$ & 41.72$^{+ 0.01}_{- 0.01}$ & 1939.03$^{+ 77}_{- 65}$ & 1196.75$^{+ 54}_{- 49}$ & 515.92 & 7.85 & 0.083\\
J075953.48+232324.2 & 0.029 & 38.30 & 43.26$^{+ 0.01}_{- 0.01}$ & 22.24$^{+ 0.02}_{- 0.02}$ & 40.74$^{+ 0.31}_{- 0.24}$ & 40.45$^{+ 0.02}_{- 0.02}$ & 571.36$^{+ 20}_{- 18}$ & 355.06$^{+ 13}_{- 13}$ & 177.95 & 5.86 & 0.005\\
J080020.98+263648.8 & 0.027 & 37.60 & 42.89$^{+ 0.01}_{- 0.01}$ & 22.15$^{+ 0.03}_{- 0.04}$ & 40.86$^{+ 0.35}_{- 0.24}$ & 40.93$^{+ 0.04}_{- 0.05}$ & 568.13$^{+ 25}_{- 19}$ & 322.17$^{+ 15}_{- 14}$ & 180.19 & 6.16 & 0.004\\
J080131.96+473616.0 & 0.157 & 40.74 & 44.26$^{+ 0.02}_{- 0.02}$ & 20.00$^{+ 0.19}$ & 44.57$^{+ 0.01}_{- 0.01}$ & 42.19$^{+ 0.01}_{- 0.01}$ & 1156.03$^{+ 130}_{- 179}$ & 721.44$^{+ 105}_{- 123}$ & 337.55 & 8.83 & 0.039\\
J080327.38+084152.2 & 0.047 & 38.16 & 43.50$^{+ 0.02}_{- 0.02}$ & 21.25$^{+ 0.11}_{- 0.13}$ & 43.13$^{+ 0.01}_{- 0.01}$ & 41.35$^{+ 0.003}_{- 0.003}$ & 535.28$^{+ 7}_{- 5}$ & 839.01$^{+ 8}_{- 7}$ & 206.66 & 7.83 & 0.014\\
      \hline
 \end{tabular}
   \end{center}
   \scriptsize
 {(1) Object ID. (2) Redshift, provided by SDSS. (3) 1.4 GHz luminosity using FIRST flux. (4) Absorption corrected 2-10 keV luminosity obtained from our X-ray spectra analysis. (5) Intrinsic Hydrogen column density. For some targets with no well-constrained $\rm N_{H}$, the values are set to the lower limit of $\rm log(N_{H})$ = 20 cm$^{-2}$. Sources with no measurements of $\rm N_{H}$ are fitted by with no intrinsic absorption component (i.e., \texttt{PL} model) in \cite{Trichas2013}. (6) Continuum luminosity at 5100\AA. (7) Total \OIII~luminosity. (8) The velocity width of the line that contains 80\% of the outflow flux. (9) The Maximal velocity of the outflow. (10) \OIII~velocity dispersion. (11) Black hole mass. (12) Eddington ratio. The entire table is available online.}
\end{table}
%=============================================================
%=============================================================

\section{Results and Discussions}\label{sec:discussion}
In this section we discuss how properties of ionized outflows may depend on X-ray, radio and accreting properties for the 183 SDSS-FIRST-X-ray-detected AGNs with ionized outflow signatures and reliable X-ray analysis at $z < 0.4$ (see Table \ref{tab:parameters} for their main parameters).

%===========
%===========

\subsection{Ionized Outflows and X-ray Properties} \label{sec:xray_properties}
\begin{figure}
\centering
\includegraphics[width=.8\linewidth]{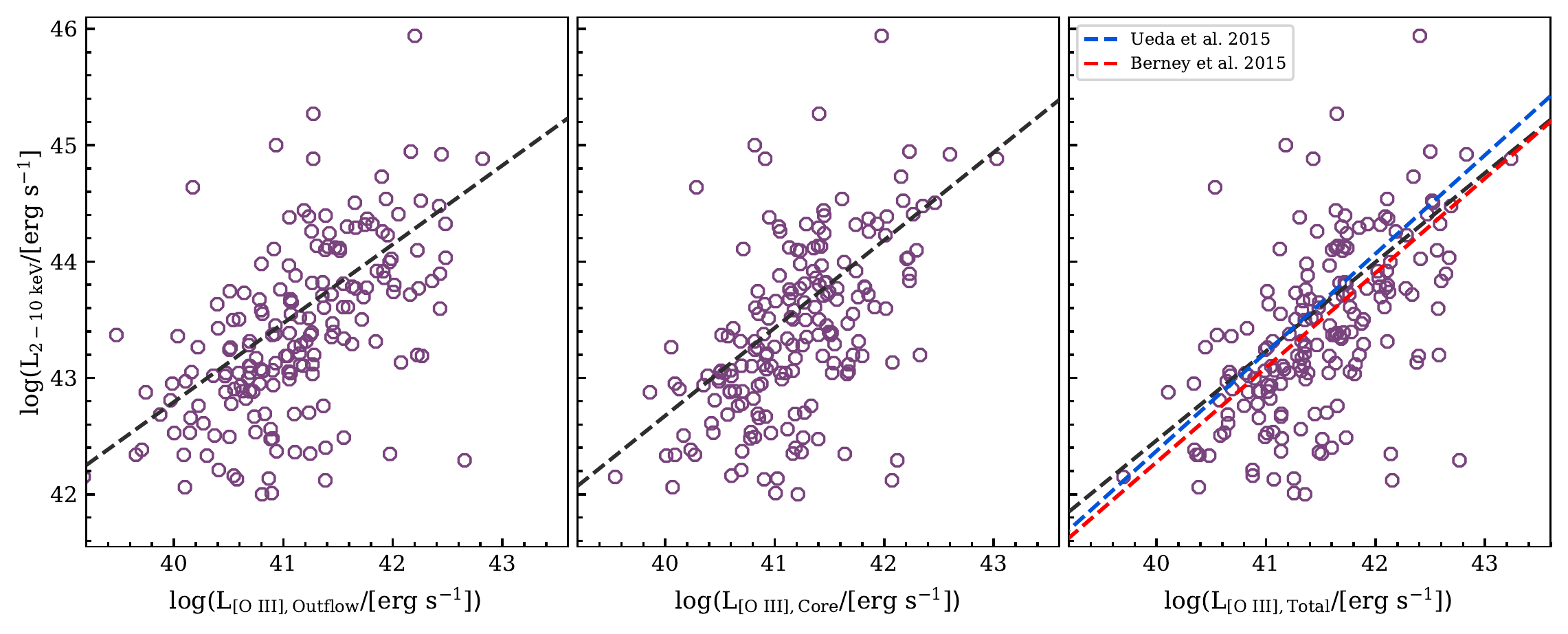}
\caption{‌The correlation between \xlum~and \OIII~luminosity derived from the outflow component (broad, $left$), core component (narrow, $middle$) and total line profile ($right$). The black dashed lines in all panels present our linear regressions. The red and blue dashed lines indicate the best-fit relation obtained by Ueda et al. \cite{Ueda2015} and Berney et al. \cite{Berney2015}, respectively.}
\label{fig:all_luminosities}
\end{figure}

As shown in previous studies \cite{Heckman2014,Duras2020}, both X-ray luminosity and the \OIII~emission-line luminosity ($\rm L_{\OIII}$) are indicators of bolometric luminosity. Therefore, it is important to fully understand the correlation between X-ray luminosity and the \OIII~luminosity for interpreting observations. In Figure \ref{fig:all_luminosities} we present the \xlum~as a function of \OIII~luminosity (from $left$ to $right$: outflow/broad component, core/narrow component and total line profile) for our sample. The Pearson $p$-value representing the probability of the null hypothesis of no $\rm L_{\OIII}$ vs. \xlum~linear correlation for all components is $\lesssim 10^{-10}$. The black dashed lines in Figure \ref{fig:all_luminosities} show the linear least-square regressions. Using the \texttt{linregress} tool of the Python \texttt{scipy} package \cite{Virtanen2020}, we find the best-fit relation of \xlum~$\propto \rm L_{\OIII, Outflow}^{0.60 \pm 0.07}$ for the outflow component and \xlum~$\propto \rm L_{\OIII, Core}^{0.66 \pm 0.08}$ for the core component. The total \OIII~luminosity correlates with X-ray luminosity as \xlum~$\propto \rm L_{\rm \OIII, Total}^{0.67 \pm 0.07}$ ($\rm \upsigma$ = 0.41 dex). Similar correlations between \xlum~and $\rm L_{\OIII}$ have also been reported in previous studies. For example, Ueda et al. \cite{Ueda2015} used a sample of hard-X-ray-selected local AGNs and found that $\rm L_{\OIII} \propto~\xlum^{1.18 \pm 0.07}$ (i.e., \xlum~$\propto \rm L_{\OIII}^{0.85}$). Consistently, \cite{Berney2015} reported that $\rm L_{\OIII}$ is proportional to \xlum$^{1.23 \pm 0.05}$ (i.e., \xlum~$\propto \rm L_{\OIII}^{0.81}$), although with large scatter ($\rm \upsigma$ = 0.61 dex), for a sample of 321 nearby ($z \simeq$ 0.04) AGNs selected above 10 keV using \swift~BAT. The blue and red dashed lines in the right panel of Figure \ref{fig:all_luminosities} present the correlations obtained in Ueda et al. \cite{Ueda2015} and Berney et al. \cite{Berney2015}, respectively.\

Perna et al. \cite{Perna2017} observed a positive trend of non-parametric outflow velocity with the unobscured 2-10 keV luminosity for a large sample ($\sim$ 550) of AGNs at $z < 0.8$ over 5 order of magnitudes and found that the most luminous AGNs have higher incidence of extreme ionized outflows. A similar trend was also observed for higher-redshift ($z \simeq$ 0.6-1.7) AGNs \cite{Harrison2016}. Here, we explore the correlation between ionized outflow velocities, derived from different tracers, and X-ray properties of our AGNs while controling radio activity. Therefore, we divide our sample into two subsamples based on radio luminosity:

\begin{itemize}
    \item Radio Luminous (RL) AGNs: \lrad~$> 10^{23}\,\rm \whz$ 
    \item Non Radio Luminous (NRL) AGNs: \lrad~$< 10^{23}\,\rm \whz$ 
\end{itemize}

\begin{figure}
\centering
\includegraphics[width=.6\linewidth]{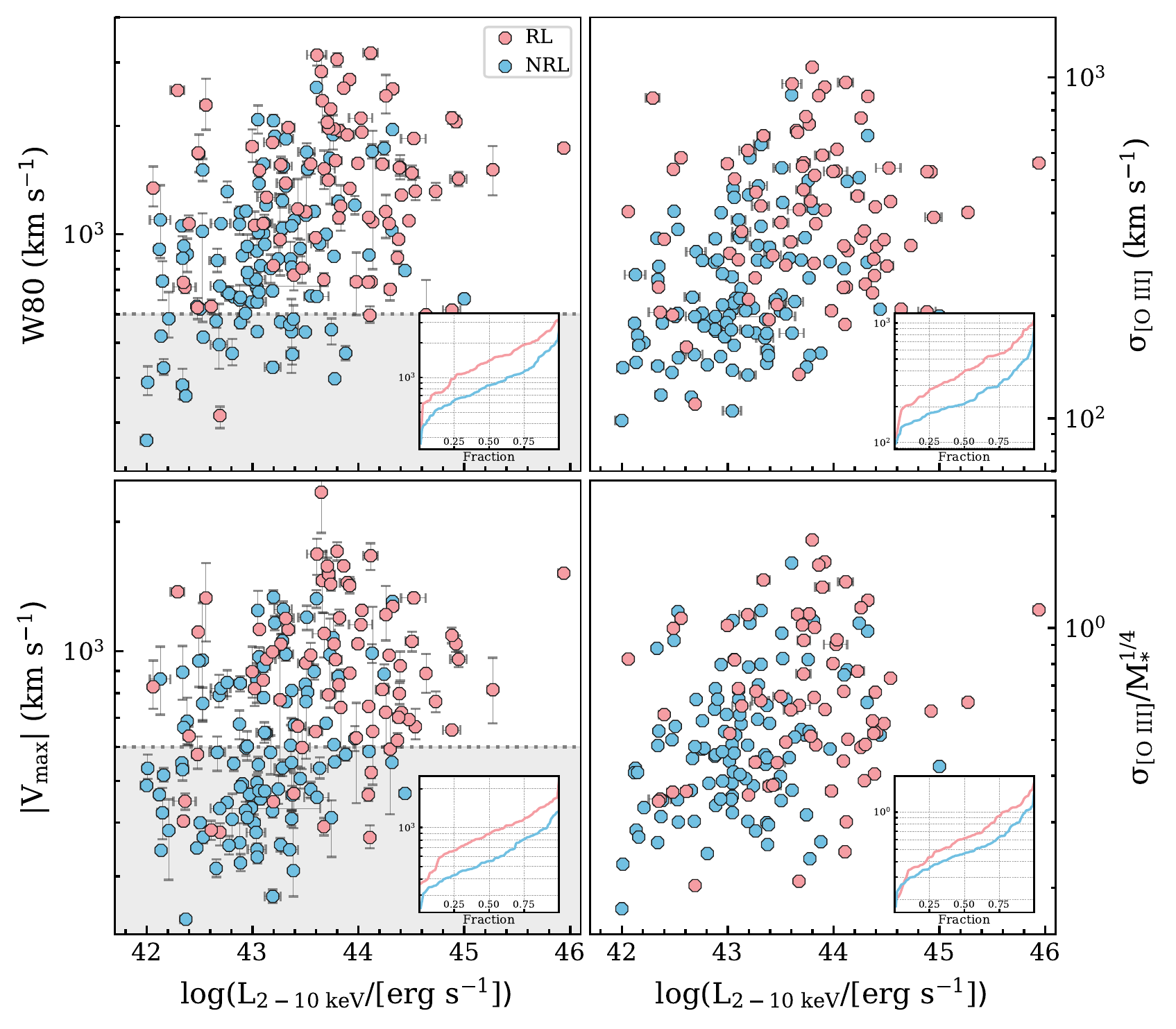}
\caption{The distributions of the maximal velocity ($lower\ left$), velocity width ($upper\ left$), \OIII~velocity dispersion ($upper\ right$), and normalized \OIII~velocity dispersion ($lower\ right$), versus \xlum. The red and blue circles exhibit the RL and NRL AGNs, respectively. The dotted lines and shaded regions indicate weak ionized outflows. {In each panel, the inset plot compares the cumulative fraction between the RL and NRL AGNs.}}
\label{fig:Lx_velocity}
\end{figure}

The lower and upper left panels of Figure \ref{fig:Lx_velocity} show the distributions of $|$V$_{\rm{max}}|$ and W80 (see Equation \ref{eq:v_max}) with \xlum, respectively. ‌‌In these panels, the shaded regions below the outflow velocity of 600 \kms~(see dotted lines) show weak ionized outflows \cite{Harrison2016}. Both W80 and V$_{\rm{max}}$ increase with 2-10 keV luminosity ($p$-value $<$ 0.000). It implies that the most X-ray luminous targets (associated with the most RL AGNs) have highest-velocity ionized outflows, consistent with the findings in Perna et al. \cite{Perna2017} and Harrison et al. \cite{Harrison2016}. To include the contributions from both the gravitational and non-gravitational \OIII~components, we calculate the the first and second moments of the total \OIII~profile from Equations 1 and 2 in Woo et al. \cite{Woo2016} using measurements which are available from BADASS output individually for broad and narrow components. Therefore, the first moment of the line profile (flux-weighted center) and the second moment (\OIII~velocity dispersion) are calculated respectively as

\begin{equation}\label{eq:first_mom}
    \rm \uplambda_{\rm{0}}= \frac{F_{\rm{B}} \uplambda_{\rm{B}} + F_{\rm{N}} \uplambda_{\rm{N}}}{F_{\rm{B}} + F_{\rm{N}}}, and
\end{equation}

\begin{equation}\label{eq:second_mom}
    \rm \upsigma^{2}_{\rm{\OIII}}= \frac{F_{\rm{B}} (\upsigma^{2}_{\rm{B}} + \uplambda^{2}_{\rm{B}}) +F_{\rm{N}} (\upsigma^{2}_{\rm{N}} + \uplambda^{2}_{\rm{N}})}{F_{\rm{B}} + F_{\rm{N}}} - \uplambda^{2}_{\rm{0}},
\end{equation}

where $\rm F_{B}$ ($\rm F_{N}$), $\lambda_{\rm B}$ ($\lambda_{\rm N}$) and $\rm \upsigma_{\rm B}$ ($\rm \upsigma_{\rm N}$) are flux, center and velocity dispersion of broad (narrow) component, respectively. In the upper right panel of Figure \ref{fig:Lx_velocity}, we show the distribution of $\rm \upsigma_{\OIII}$ against \xlum. As expected, \OIII~velocity dispersion increases with 2-10 keV X-ray luminosity. It is indicative of photoionization of gas in narrow line region (NLR) by X-ray emission from AGN. An alternative explanation is that there is a tight link between X-ray luminosity and radio luminosity \cite{Ballo2012,Liao2020,Chang2021}, and that, as Woo et al. \cite{Woo2016} discussed, more radio luminous AGNs (hosted in more massive galaxies) tend to have broader \OIII~profiles due to large gravitational potential. Therefore, we may observe that more X-ray luminous AGNs have higher \OIII~velocity dispersion.

We also try to examine the correlation between X-ray luminosity and \OIII~velocity dispersion normalized by the stellar velocity dispersion to account for motion of ionized gas due to the galactic gravitational field \cite{Karouzos2016,Luo2021}. Based on the our FWHM (H$\beta$) measurements and image cutouts of SDSS images, a large fraction of our sample consist of Type 1 AGNs (FWHM (H$\beta$) $>$ 1000 \kms) viewed as face-on galaxies. Since the estimation of stellar velocity dispersion in bright Type 1 AGNs is more complicated, we use stellar mass ($\rm M_{*}$) as a surrogate of the host-galaxy gravitational potential. We obtain stellar mass measurements for our sources (except 9 targets) from the MPA-JHU Catalogue \cite{Abazajian2009}. In the lower right panel of Figure \ref{fig:Lx_velocity} we show the distribution of normalized \OIII~velocity dispersion ($\rm \upsigma_{\OIII}$/$\rm M_{*}^{1/4}$) versus \xlum. We find that the normalized \OIII~velocity dispersion increases with X-ray luminosity. However, to constrain the correlation between \xlum~ and $\rm \upsigma_{\OIII}$/$\rm M_{*}^{1/4}$ reliably, a large sample is required in further studies. We find that the normalized \OIII~velocity dispersion increases with radio luminosity, however scattering is slightly larger. {We also compare the cumulative fraction of non-parametric outflow velocities between the RL and NRL AGNs as inset plots in Figure \ref{fig:Lx_velocity}. It is clear that a higher fraction of the RL targets are associated with broader \OIII~}{emission lines (likely due to the stronger gravitational potential in the RL AGNs). This difference trends to be weaker when the \OIII~}{velocity dispersion is normalized with stellar mass. It is in agreement with the findings of Woo et al. \cite{Woo2016} and Rakshit and Woo \cite{Rakshit2018} who found that the increasing trend of the \OIII~}{velocity dispersion as a function of radio luminosity is weak once the \OIII~}{velocity dispersion is normalized by stellar velocity dispersion. It suggests that it is important to correct the host-galaxy effect when we examine the kinematics of ionized gas outflows.}\

\begin{figure}
\centering
\includegraphics[width=.6\linewidth]{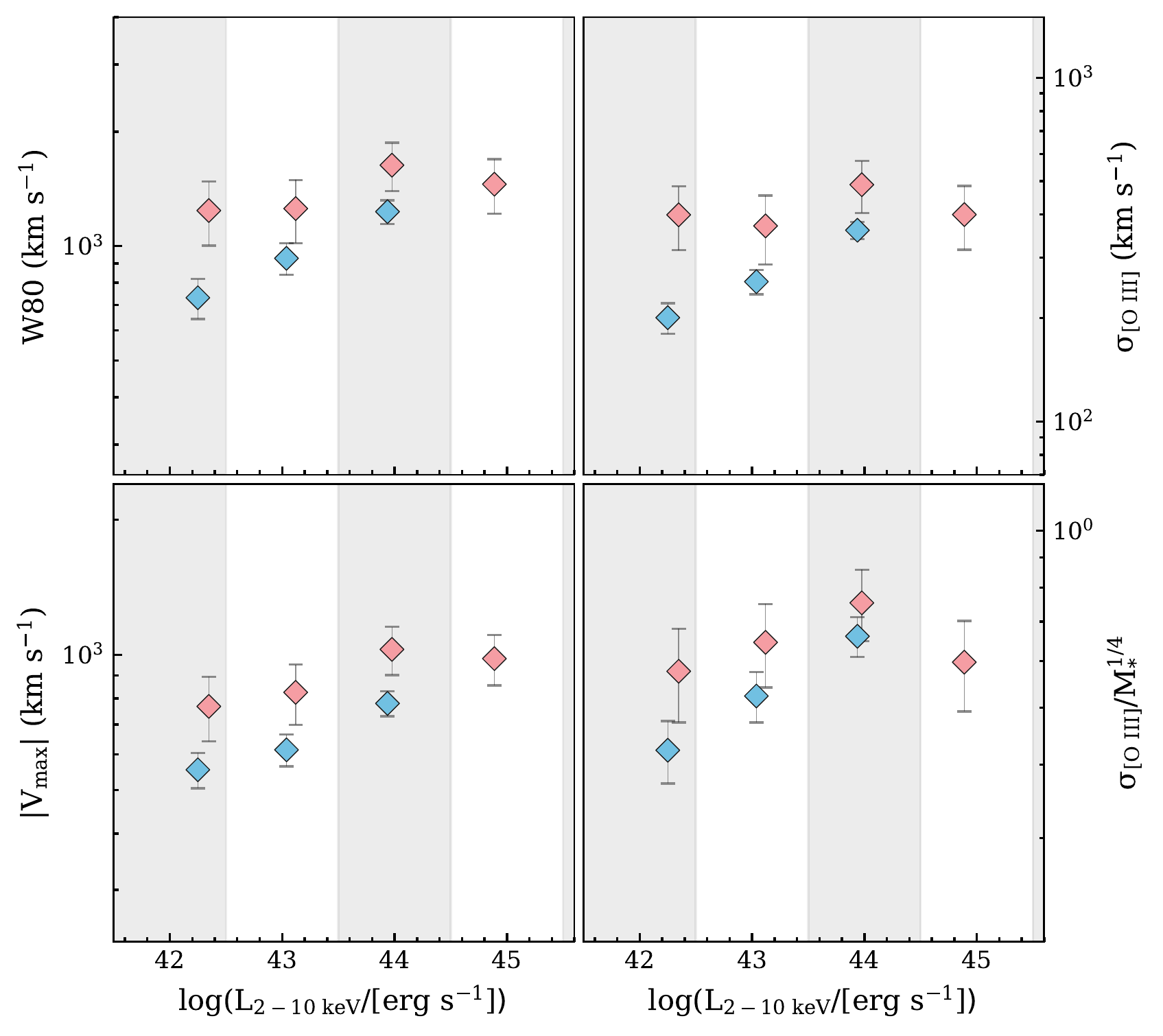}
\caption{‌Same as Figure \ref{fig:Lx_velocity}, but  \xlum~is in bins of 1 dex. Diamonds present the mean values in each bin for each sample. The error bars show the Root-Mean-Square error (RMSE) in each bin.}
\label{fig:Lx_velocity_binned}
\end{figure}

The prevalence of ionized outflows with $|\rm V_{max}|$ (W80) $>$ 600 \kms~increases from 43\% (78\%) in the NRL sample to 83\% (96\%) in the RL samples. Since targets with higher radio luminosity are preferentially associated with X-ray bright AGNs, is not clear whether the higher fraction of extreme ionized outflows in more radio luminous AGNs is attributed to the radio activity or not. Hence, to determine the role of radio activity, we fix X-ray luminosity ($\rm \upDelta L_{2-10\,keV}$ is 1 dex) and compare the mean values of non-parametric tracers of ionized outflows in the RL and NRL AGNs in Figure \ref{fig:Lx_velocity_binned}. We observe that for a given X-ray luminosity, the RL AGNs have higher mean outflow velocity in almost all bins. These results suggest that radio jets may contribute in driving high-velocity outflows. We investigate the role of radio activity in more details in Section \ref{sec:radio_properties}.\

\begin{figure}
\centering
\includegraphics[width=.55\linewidth]{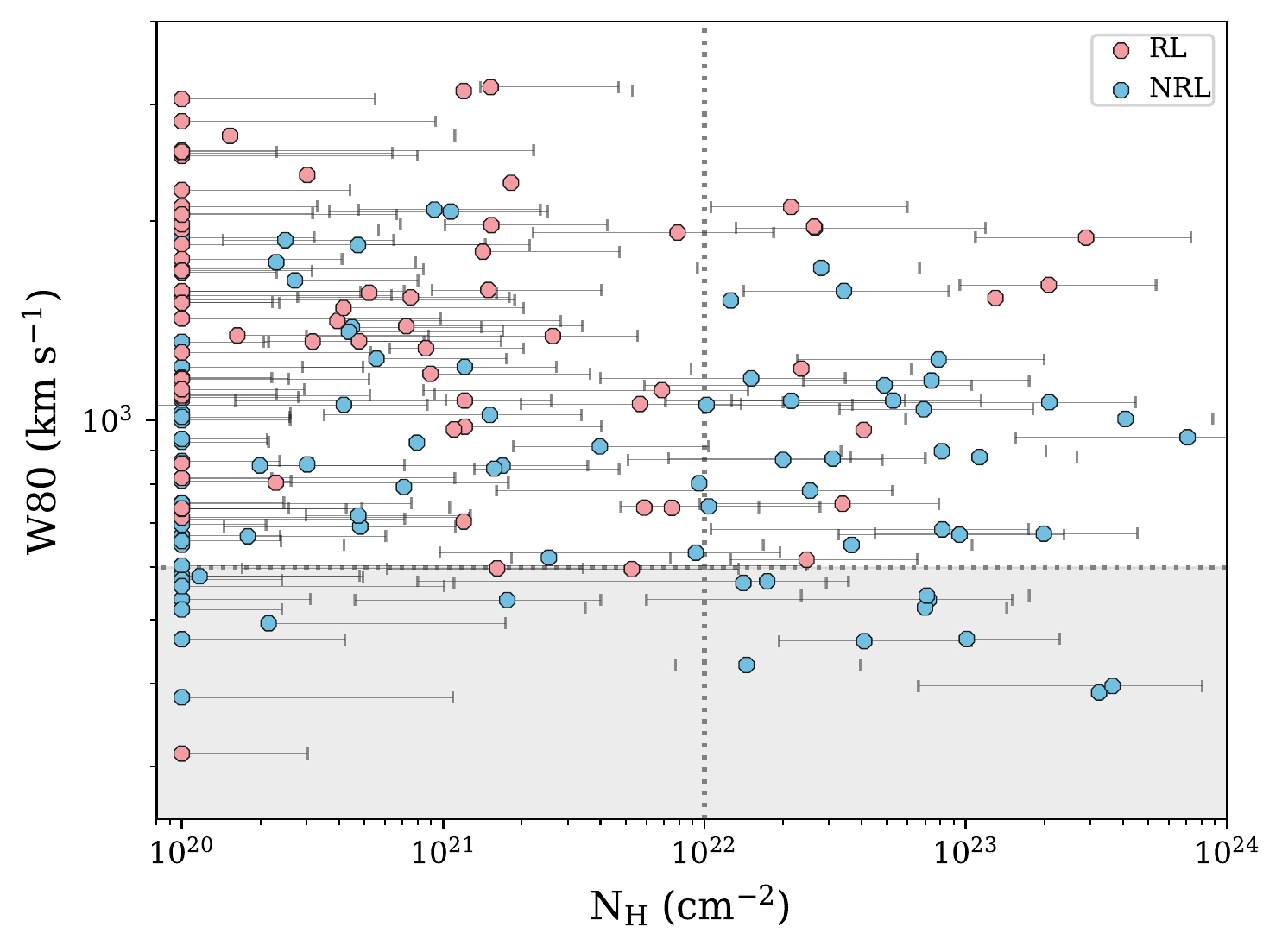}
\caption{‌The \OIII~velocity width (W80) versus hydrogen column density ($\rm N_{\rm H}$). The vertical dotted line separates X-ray obscured ($\rm N_{\rm H} > 10^{22} cm^{-2}$) and unobscured ($\rm N_{\rm H} < 10^{22} cm^{-2}$) AGNs.}
\label{fig:nh_w80}
\end{figure}

Some studies suggest that AGN activity is triggered by galaxy merger events through redistributing of angular momentum of the gas and transporting the gas inward. In this scenario, AGNs follow an evolutionary sequence from heavily obscured AGNs, where they are hosted in massive star-forming galaxies and produce similar kinematic signatures as outflows, to unobscured AGNs \cite{Hopkins2008}. Therefore, it might be expected that obscured AGNs may be more associated with high-velocity ionized outflows.
The high prevalence of extreme outflows in dust-obscured Ultra Luminous Infrared Galaxies (ULIRGs) frequently found in interacting and merging galaxies may support this idea. Therefore, we explore the correlation between the \OIII~emission-line velocity width and obscuring column density (N$_{\rm H}$) estimated from our X-ray spectral analysis. We define obscured (unobscured) AGNs with a column density greater (smaller) than $\rm N_{H} = 10^{22} cm^{-2}$. Figure \ref{fig:nh_w80} shows that there is no significant difference between velocity widths of obscured AGNs and that of unobscured targets. Indeed, the prevalence of ionized outflows with W80 $>$ 600 \kms~in unobscured AGNs is slightly higher than those in obscured AGNs by a factor of $\sim$ 1.2. However, this might be due to the fact that X-ray obscured AGNs with high column density are preferentially Type 2 AGNs in which  the projection and orientation effect impacts on detection of ionized outflow signatures \cite{Rakshit2018}. Therefore, our findings suggest that obscured AGNs are not preferably associated with higher-velocity outflows (see also \cite{Harrison2016,Rojas2020}).

\subsection{Ionized Outflows and Radio Properties} \label{sec:radio_properties}
The connection between outflowing gas and radio activity is still controversial. While some individual studies propose that ionized outflows can be driven through the interaction of the radio jet with the ISM {\cite{Couto2020,Girdhar2022}}, others discuss that the connection between radio jet and outflowing gas depends on their morphology and position angle \cite{Le2017}. {Some simulation studies (see Mukherjee et al. \cite{Mukherjee2021} for a review) suggest that the distribution of gas in the host-galaxy can be significantly impacted by relativistic radio jets, however, the resultant velocities might be insufficient to unbind an important portion of the gas.} In this section, we explore the connection between radio activity and highest-velocity ionized outflows in our sample. 
Similar to Section \ref{sec:xray_properties}, while we investigate the role of radio emission, we control X-ray activity. Some previous studies reported that the fraction of absorbed AGNs change against the luminosity around $\rm log(L_{X}) \sim 43.5$ \ergs~\cite{Burlon2011,Ueda2011}. It implies that sources with $\rm log(L_{X}) > 43.5$ \ergs~are less absorbed and more luminous. Therefore, we divide our sample into two subsamples based on this value as the threshold. 

\begin{itemize}
    \item X-ray Luminous (XL) AGNs: \xlum~$> 10^{43.5}\,\rm \ergs$ 
    \item Non X-ray Luminous (NXL) AGNs: \xlum~$< 10^{43.5}\,\rm \ergs$
\end{itemize}

\begin{figure}
\centering
\includegraphics[width=.6\linewidth]{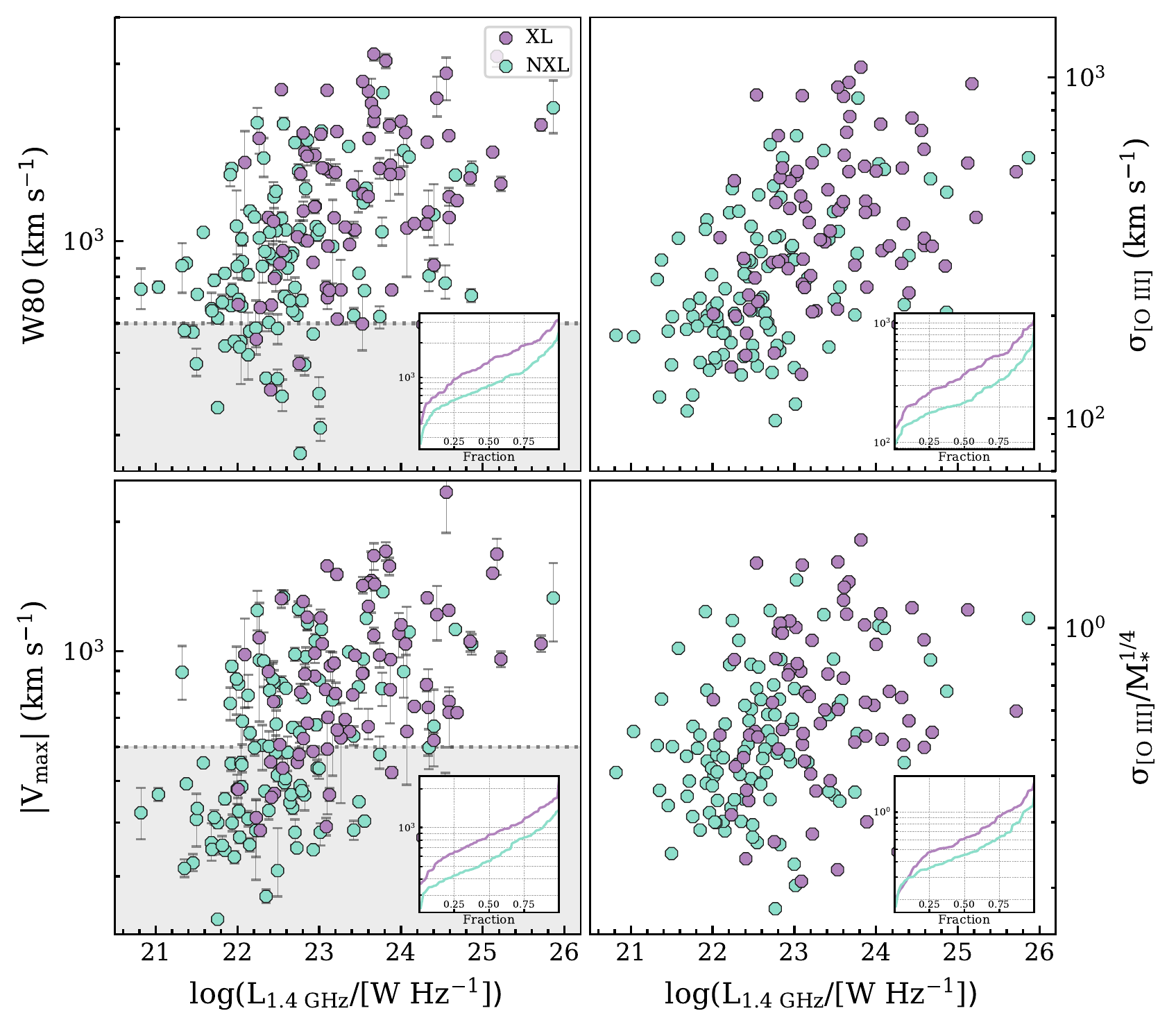}
\caption{‌Same as Figure \ref{fig:Lx_velocity}, but X-axes show radio luminosity. The purple and green circles exhibit the XL and NXL AGNs, respectively. The dotted lines and shaded regions indicate weak ionized outflows. {In each panel, the inset plot compares the cumulative fraction between the RL and NRL AGNs.}}
\label{fig:LR_velocity}
\end{figure}

We note that the above threshold is almost same as the mean and median values of X-ray luminosity in our sample (i.e, log(\xlum)~$\sim$ 43.4 \ergs). Our results are not sensitive to the selection of either log(\xlum)~$\sim$ 43.4 or log(\xlum)~$\sim$ 43.5 \ergs. Figure \ref{fig:LR_velocity} presents the distributions of four non-parametric outflow quantities as a function of 1.4 GHz luminosity. All shaded regions are characterized same as Figure \ref{fig:Lx_velocity}. It is obvious that $|$V$_{\rm{max}}|$ (lower left panel), W80 (upper left panel) and $\rm \upsigma_{\OIII}$(upper right panel) tightly correlate with \lrad. Woo et al. \cite{Woo2016} discussed that the apparent correlation between \OIII~velocity dispersion and radio luminosity is due to the fact that radio luminosity increases with stellar mass (and stellar velocity dispersion) and accordingly, in more massive galaxies \OIII~emission-line profiles are broader due to the larger gravitational potential. Therefore, more radio luminous galaxies are preferentially associated with more broader \OIII~profiles.  Here, we follow Woo et al. \cite{Woo2017} to correct the effect of the host-galaxy, and examine the correlation between \OIII~velocity dispersion normalized by stellar mass ($\rm \upsigma_{\OIII}$/$\rm M_{*}^{1/4}$) and radio luminosity in the lower right panel of Figure \ref{fig:LR_velocity}. These results may suggest the hypothesis that radio activity is capable of contributing in driving high velocity ionized outflows in our sample. {Similar to the fractional comparison done between the RL and NRL AGNs in Figure \ref{fig:Lx_velocity}, we compare the cumulative fractions of outflow velocities between the XL and NXL AGNs (see inset plots in Figure \ref{fig:LR_velocity}). The results suggest that the XL targets tend to exhibit higher outflow velocities than NXL sources, likely due to the sweeping-up of ambient material by their stronger radiation pressure. This difference is weaker when we normalize the \OIII}{~velocity dispersion.}\

\begin{figure}
\centering
\includegraphics[width=.6\linewidth]{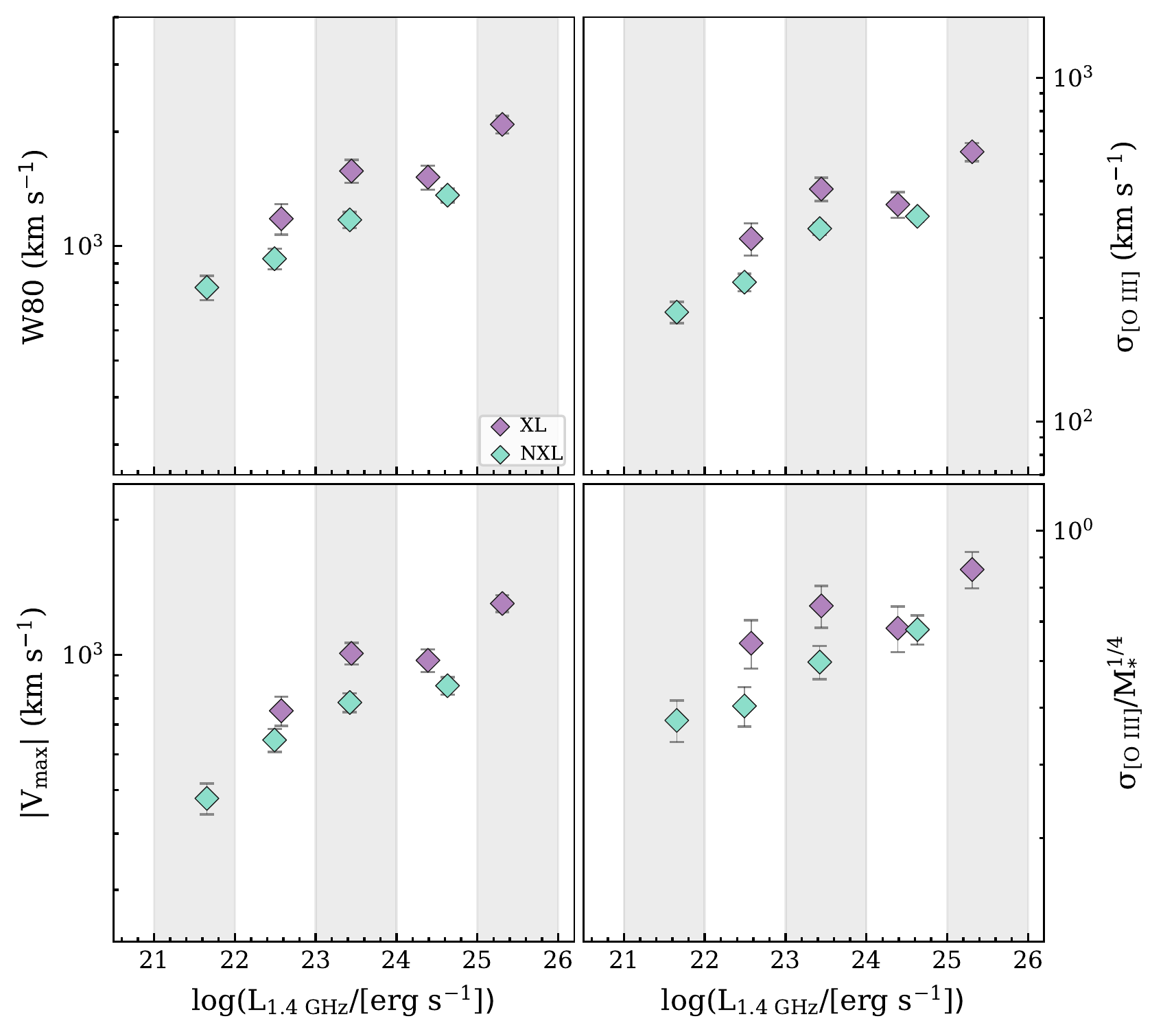}
\caption{‌‌Same as Figure \ref{fig:LR_velocity}, but  \lrad~is in bins of 1 dex. Diamonds present the mean values in each bin for each sample. The error bars show Root-Mean-Square error (RMSE) in each bin.}
\label{fig:LR_velocity_binned}
\end{figure}

For further verification of the above results, we explore the role of X-ray activity in driving ionized outflows in sources with the same radio luminosity. Therefore, we fix 1.4 GHz radio luminosity ($\Delta$\lrad~is 1 dex) and reproduce Figure \ref{fig:LR_velocity} using the mean values of outflow velocity for the XL and NXL samples in Figure \ref{fig:LR_velocity_binned}. For a given radio luminosity, the XL AGNs have a higher mean outflow velocity than the NXL AGNs. It indicates that X-ray luminosity is also {connected} to extreme ionized outflows, comparable to radio luminosity.\

\begin{figure}
\centering
\includegraphics[width=.45\linewidth]{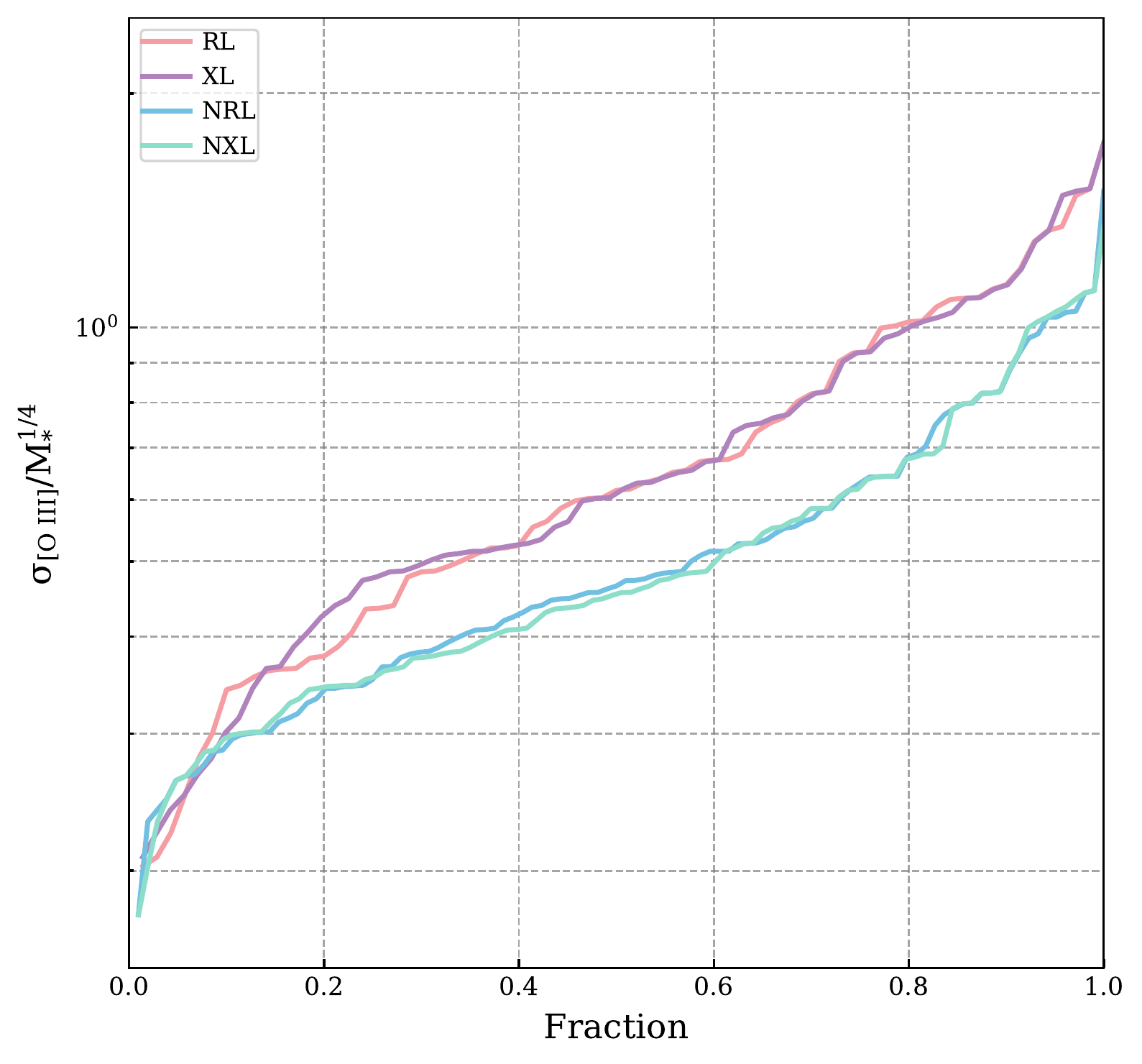}
\caption{‌{The cumulative fractions of the normalized \OIII}{~velocity dispersion of our classified subsamples. Colors are identical to Figures \ref{fig:Lx_velocity} and \ref{fig:LR_velocity}.}}
\label{fig:fraction}
\end{figure}

We also perform the partial-correlation (PCOR) test using \texttt{pcor} in $R$ to statistically examine the above correlations. One parametric test (Pearson) and two non-parametric tests (Spearman and Kendall) are available. Therefore, we check the correlation between X-ray (radio) luminosity and ionized gas kinematics while controlling for the radio (X-ray) luminosity. Table \ref{tab:pcor} presents the $p$-values for each method. The results are in agreement with Figure \ref{fig:Lx_velocity_binned} and \ref{fig:LR_velocity_binned} and suggest that ionized outflows correlate slightly stronger with radio luminosity than X-ray luminosity. However, we note that due to the small sample size, these statistics may not be very certain. {As a further check, Figure \ref{fig:fraction} presents the comparison of the cumulative fraction of the normalized \OIII}{~velocity dispersion between the XL (NXL) and RL (NRL) subsamples. It is evident that there is no significant difference in prevalence of high-velocity ionized outflows between the XL (NXL) and RL (NRL) targets.} 

\begin{figure*}
\centering
\includegraphics[width=.9\linewidth]{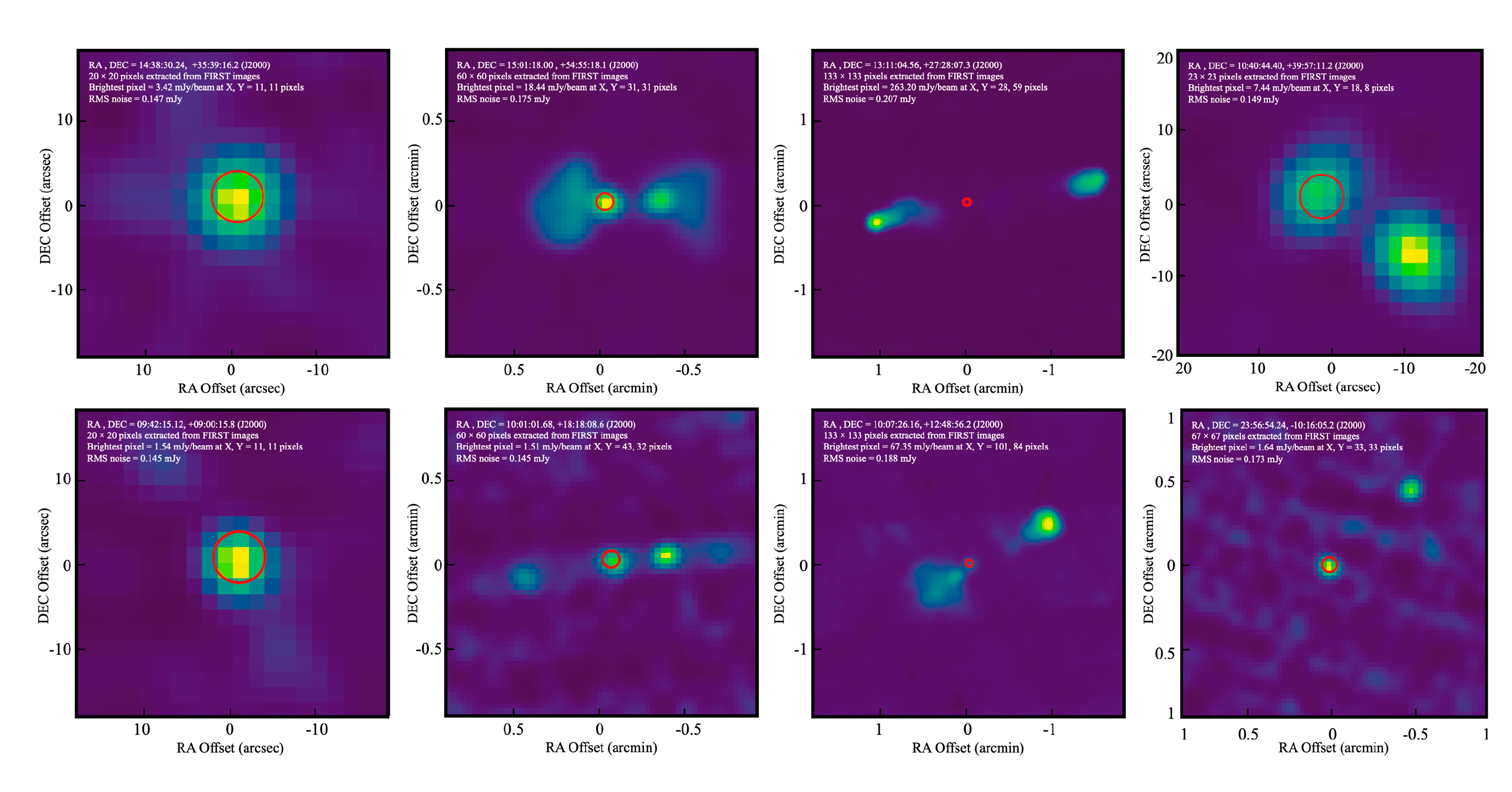}
\caption{‌Examples of radio morphology classified by the FIRST Classifier \cite{Alhassan2018}. From $left$ to $right$: Compact, FRI, FRII and BENT. The red circle in each panel shows the SDSS fibre size.}
\label{fig:morphology}
\end{figure*}

Using intermediate-resolution observations of 14 radio galaxies, Holt et al. \cite{Holt2008} concluded that compact radio sources (radio structure confined to a $<$ 10 kpc region) harbour more extreme ionized outflows than their extended counterparts. This result was recently confirmed by Molyneux et al. \cite{Molyneux2019} using a large sample of 2,922 radio sources at $z < 0.2$. These studies may suggest an ionized outflow- radio connection in compact radio sources. Hence, we study the radio morphology of our sample. We classify our sample using the Deep Convolutional Neural Networks model applied on FIRST images (The FIRST Classifier; Alhassan et al. \cite{Alhassan2018}) into compact (unresolved sources which have a single non-diffuse component) and extended (FRI, FRII and BENT) radio AGNs. The overall accuracy of morphology classification using this technique for their parent sample was $\sim$ 97\% \cite{Alhassan2018}.  We find that our sample consists of $\sim$ 78\% (141/183) compact, $\sim$ 18\%  (34/183) either FRI or FRII and  $\sim$ 3\% (5/183) BENT radio AGNs. Three sources remain morphologically unclassified. There are only 13 radio sources (including 5 compact and 8 extended) in our sample that their highest probability of classification is less than 90\%. Indeed, about 90\% (162/183) of our targets are classified with probability more than 99\%. Figure \ref{fig:morphology} shows some examples of our morphology classification. Moreover, we also examine the distribution of the major-axis sizes of our sample obtained from two-dimensional Gaussian fits (deconvolved for the beam). The major-axis size is less than 3 arcsec ($\sim$ 5 kpc at the median redshift of our sample, $z$ = 0.1) in 64\% (115/183) and less than 5 arcsec ($\sim$ 9 kpc at $z$ = 0.1) in 86\% (156/183) of our targets. {According to these results, our sample mainly consists of compact radio AGNs whose radio emission size is comparable to that of ionized outflows ($\sim$ 1-10 kpc \cite{Greene2011,Fischer2013}). Therefore, we conclude that low-power compact radio jets can accelerate high-velocity ionized gas outflows, in accordance with some previous studies.}\

\begin{table}
\begin{center}
\caption{$p$-values of Partial Correlation Analyses.}
\label{tab:pcor}
\begin{tabular}{cccc}\hline\hline
Relation & Pearson & Spearman & Kendall \\
\hline
\xlum-$|\rm V_{max}|$  &
		$0.019$   &
		$0.02$ &  
		$10^{-3}$ \\
\xlum-W80&  $10^{-3}$  &
		 $10^{-3}$   &
		 $10^{-4}$\\
\xlum-$\sigma_{\OIII}$  &
		 $10^{-3}$  &
		 $10^{-3}$   &
		 $10^{-4}$\\
\xlum-$\sigma_{\OIII}/M_{*}^{1/4}$&
		 $0.06$  &
		 $0.05$   &
		 $10^{-3}$\\ 		 
		  \hline
\lrad-$|\rm V_{max}|$  &
		$10^{-6}$   &
		$10^{-7}$ &  
		$10^{-8}$ \\
\lrad-W80&  $10^{-5}$  &
		 $10^{-6}$   &
		 $10^{-6}$\\
\lrad-$\sigma_{\OIII}$&
		 $10^{-6}$  &
		 $10^{-6}$   &
		 $10^{-8}$\\
\lrad-$\sigma_{\OIII}/M_{*}^{1/4}$&
		 $10^{-3}$  &
		 $10^{-3}$   &
		 $10^{-3}$\ \\ \hline
\end{tabular}
\end{center}
%{\sc Note.} --- a: the $p$-value is too small for
%	        significance calculation.\\
\end{table}

\subsection{Ionized Outflows and Eddington ratios} \label{sec:eddington_ratio}
\begin{figure}
\centering
\includegraphics[width=.6\linewidth]{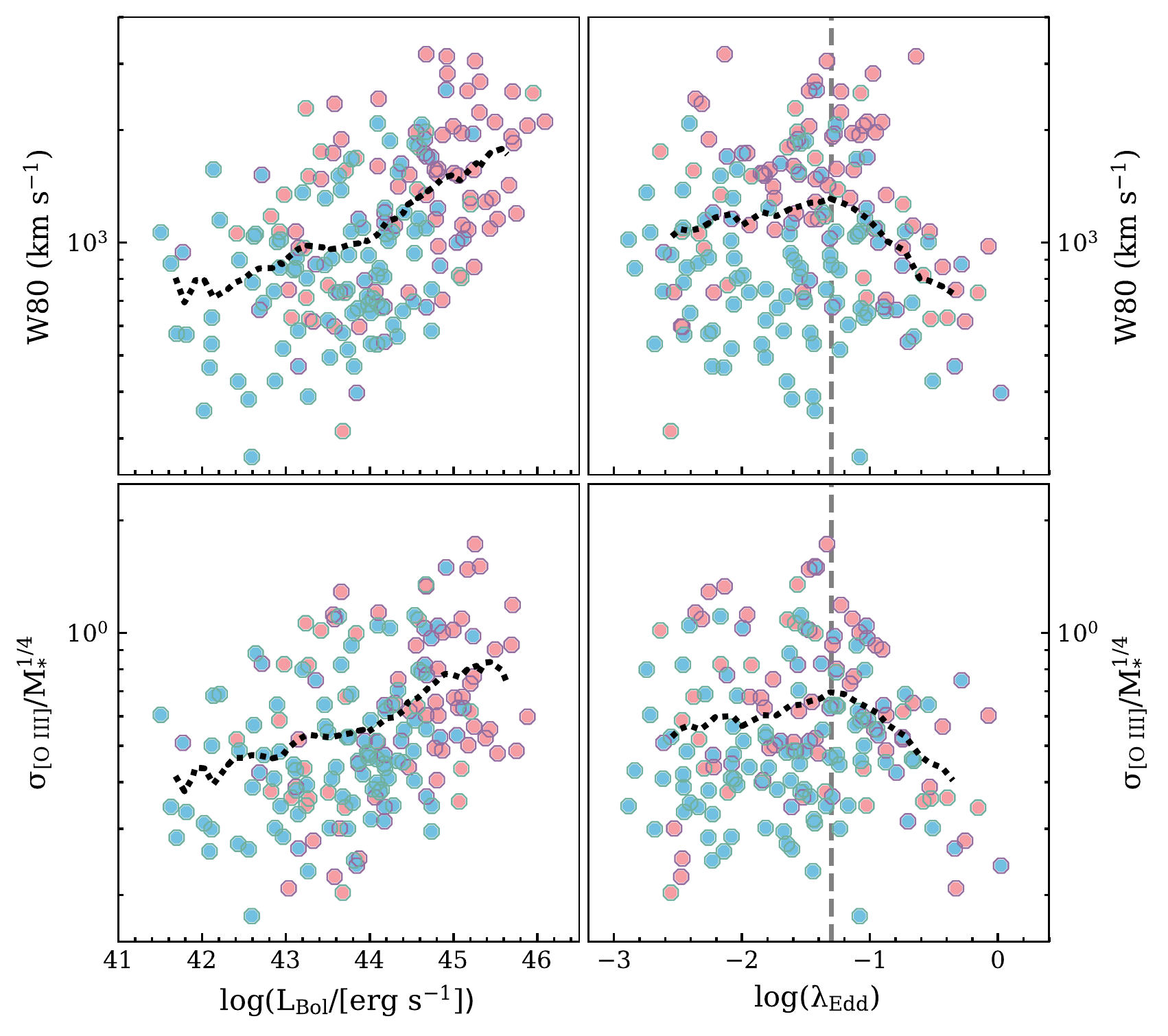}
\caption{‌The distribution of W80 ($top$) and normalized \OIII~velocity dispersion ($bottom$) as a function of bolometric luminosity ($left$) and Eddington ratio ($right$). The black dotted lines illustrate the running mean in each panel. The grey dashed lines in the right panels indicate $\rm log (\uplambda_{Edd}) = -1.3$. The Colors are identical to that in Figure \ref{fig:Lx_velocity} and \ref{fig:LR_velocity}.}
\label{fig:Eddinton_ratio}
\end{figure}

Previous studies of ionized gas outflows have concluded that the AGN Eddington ratio ($\rm \uplambda_{Edd}$) plays a key role in dictating the kinematics of the non-gravitational component. For example, Rakshit and Woo \cite{Rakshit2018} and Woo et al. \cite{Woo2016} reported a steep increase of both the outflow velocity (i.e., \OIII~velocity dispersion normalized by stellar velocity dispersion) and outflow fraction (i.e., the fraction of AGNs that show a wing component in the \OIII~line profile) with Eddington ratio for a large sample of Type 1 and Type 2 AGNs, respectively. In this context, Rojas et al. \cite{Rojas2020} found a dichotomy between Type 1 and Type 2 AGNs as the outflow fraction for Type 1+1.9 AGNs generally decreases or remains flat with accretion rate, while outflow incidence for Type 2 AGNs increases dramatically. Then, the outflow fraction in both type 1+1.9 and type 2 AGN starts to decrease at $\rm log(\uplambda_{Edd}) \gtrsim -1.2$. They interpreted their results in the framework of covering factor-Eddington ratio dependence (see their Section 5.4 for more details). Therefore, in this section we study properties of Eddington ratio in our sample. In Section \ref{sec:others}, we describe the estimation of Eddington ratio for our sample. The Eddington ratio of our sample spans over $\rm -3 \lesssim log(\uplambda_{Edd}) \lesssim 0$ with a median of -1.53. It suggests that our sample is composed of the low-to-moderately accreting sources. In Figure \ref{fig:Eddinton_ratio} we show the distributions of line velocity width (W80) (top row) and the normalized \OIII~velocity dispersion (bottom row) as a function of bolometric luminosity (left panels) and Eddington ratio (right panels). In agreement with previous studies, we find that outflow velocity (traced by either W80 or $\rm \upsigma_{\OIII}$/$\rm M_{*}^{1/4}$) increases with bolometric luminosity. Interestingly, we find that outflow velocity increases with Eddington ratio when $(\uplambda_{Edd}) \lesssim -1.3$ and then it decreases. {From the theoretical perspective, Ishibashi et al. \cite{Ishibashi2018} demonstrated that the observed outflow energetics can be adequately explained by AGN radiation pressure on the host-galaxy dust. In this scenario, the radiation field of an AGN can be strongly coupled with the surrounding dusty gas resulting AGN-driven outflows. Therefore, there is an increasing trend of the outflow velocity as a function of the Eddington ratio.} 

When the Eddington ratio hits the highest values (log$(\uplambda_{Edd}) \gtrsim -1.3$), most of the circumnuclear materials have already been expelled from the vicinity of the black hole and outflows are not very strong. However, due to the lack of super-Eddington accreting sources in our sample, we are not in a position to examine the expelling effect in the super-Eddington regime. We also note that the fractional number of the XL (RL) and NXL (NRL) before and after log$(\uplambda_{Edd}) \gtrsim -1.3$ is very close. Therefore, our results are not affected by distribution of our sources. \

{According to our systematic study in which we assess the role of radio luminosity and X-ray luminosity in driving high-velocity ionized outflows, no clear preference between X-ray luminosity and radio luminosity in driving high-velocity ionized outflows is found. However, our sample may suffer from incompleteness effects especially in terms of X-ray luminosity. Furthermore, the majority of our targets are radio compact AGNs in which the interaction with radio jets may play a role in driving more extreme ionized gas kinematics. Therefore, large samples of representative AGNs in future would be desired to improve the sample completeness and the reliability of correlation analyses.}

Given the data presented in this work, we are not able to determine the exact origin (if there were indeed a single origin) of outflows in our sample. One scenario is that they may be driven by radiation pressure in the nuclear region (AGN-induced), then due to the contribution of radio jets, they are accelerated up to high velocities. Alternatively, the interaction of radio jet with ISM may launch and accelerate ionized outflows (jet-induced). However, the evidence for the latter is limited to individual sources and there is still no enough observational support for large samples of outflows. Since our sample generally includes strong ionized outflows, we conclude that our results are in favour of the former.

\section{Conclusions} \label{sec:conclusions}
In this study we uncover the relation of ionized gas kinematics with X-ray and radio properties as well as Eddington ratio in a sample of 348 SDSS-FIRST-X-ray detected AGNs at z $<$ 0.4. To obtain X-ray properties of our sample, we utilized X-ray observations from \xmm, \swift~ and \chandra~surveys. Our main results are summarized
as follows:

\begin{enumerate}

\item
Approximately 63\% (220/348) of our targets exhibit a non-galactic broad component in their \OIII~profile as an indicative of ionized outflows. It suggest the high prevalence of ionized outflows in radio AGNs.\

\item
The non-parametric outflow velocities (i.e., velocity width, the maximal velocity of outflow and line dispersion) have positive correlations with both X-ray and radio luminosity. Moreover, the \OIII~velocity dispersion normalized by the stellar mass also correlates with both X-ray and radio luminosity. \

\item
For a given X-ray luminosity, we find that the radio-luminous sample has a higher outflow velocity in comparison with the non-radio luminous sample. Similarly, for a given radio luminosity we find that X-ray luminous AGNs have slightly higher outflow velocities than non-X-ray luminous AGNs.  We find no clear preference between X-ray luminosity and radio luminosity in driving high-velocity ionized outflows.\

\item
We find no evidence that obscured AGNs host more powerful outflows in comparison to unobscured AGNs.\

\item
Based on our estimation of black hole mass using FWHM(H$\beta$) and continuum luminosity at 5100\AA, we find that the Eddington ratio of our sample covers the interval of $-3 \lesssim$ log$(\lambda_{Edd}) \lesssim 0$ with a median of -1.58. It may imply that our sample is composed of the low-to-moderately accreting sources. When we explore the dependency of outflow velocity on Eddington {ratio}, we find that outflow velocity increases with Eddington ratio when log$(\lambda_{Edd}) \lesssim -1.3$, and when Eddington ratio become larger, outflow velocity decreases. This critical value is in agreement with previous studies that suggested when Eddington ratio reaches high values (log$(\lambda_{Edd}) \gtrsim -1.3$), most of circumnuclear material has been pushed away and outflows cannot be accelerated effectively. {However, these results may be further tested using more complete samples because our sample does not have many super-Eddington accreting sources. } \

\end{enumerate}

%%%%%%%%%%%%%%%%%%%%%%%%%%%%%%%%%%%%%%%%%%

\authorcontributions{Conceptualization, Ashraf Ayubinia; Data curation, Ashraf Ayubinia; Formal analysis, Ashraf Ayubinia; Supervision, Yongquan Xue and Jong-Hak Woo; Validation, Huynh Anh Nguyen Le, Zhicheng He, Halime Miraghaee and Xiaozhi Lin; Writing – original draft, Ashraf Ayubinia; Writing – review \& editing, Yongquan Xue, Jong-Hak Woo, Huynh Anh Nguyen Le, Zhicheng He, Halime Miraghaee and Xiaozhi Lin.}

\acknowledgments{A. A acknowledges support from China Scholarship Council for the Ph.D. Program (No. 2017SLJ021244). A. A., Y. Q. X., H. A. N. L. and X. Z. L. acknowledge support from the National Natural Science Foundation of China (NSFC-12025303, 11890693, 12003031), the K. C. Wong Education Foundation, and the science research grants from the China Manned Space Project with NO. CMS-CSST- 2021-A06. J. H. W. acknowledges the support by the Basic Science Research Program through the National Research Foundation of Korean Government (NRF-2021R1A2C3008486). H.A.N.L. acknowledges the support from the "Fundamental Research Funds for the Central Universities". Z. H. acknowledges support from the National Natural Science Foundation of China (NSFC-12222304).}

\institutionalreview{Not applicable.}

\conflictsofinterest{The authors declare no conflict of interest.} 
%%%%%%%%%%%%%%%%%%%%%%%%%%%%%%%%%%%%%%%%%%
\reftitle{References}

\end{document}